\documentclass{aa}  

\usepackage{graphicx}
\usepackage[dvipsnames]{xcolor}
\usepackage[normalem]{ulem} 
\usepackage{txfonts}
\usepackage[colorlinks=true
,urlcolor=blue
,anchorcolor=blue
,citecolor=blue
,filecolor=blue
,linkcolor=blue
,menucolor=blue
,linktocpage=true
,pdfproducer=medialab
,pdfa=true]{hyperref}

\begin{document}

   \title{Dark matter halo dynamics in 2D Vlasov Simulations:}

   \subtitle{A Self-similar approach}

   \author{Abineet Parichha
          \inst{1}\fnmsep\thanks{corresponding author}
          ,
          Stephane Colombi\inst{1}
          ,
          Shohei Saga\inst{1,2,3}
          \and
          Atsushi Taruya\inst{4,5}
          }

   \institute{$^1$ Sorbonne Universit\'e, CNRS, UMR7095, Institut d'Astrophysique de Paris, 98bis boulevard Arago, F-75014 Paris, France\\
   $^2$ Institute for Advanced Research, Nagoya University, Furo-cho Chikusa-ku, Nagoya 464-8601, Japan\\
   $^3$ Kobayashi-Maskawa Institute for the Origin of Particles and the Universe, Nagoya University, Chikusa-ku, Nagoya, 464-8602, Japan\\
   $^4$ Center for Gravitational Physics and Quantum Information, Yukawa Institute for Theoretical Physics, Kyoto University, Kyoto
   606-8502, Japan\\
   $^5$ Kavli Institute for the Physics and Mathematics of the Universe (WPI), Todai institute for Advanced Study, University of Tokyo,
   Kashiwa, Chiba 277-8568, Japan\\
              \email{abineet.parichha@iap.fr}
             \thanks{author email}
             }

   \date{\today}

% \abstract{}{}{}{}{} 
% 5 {} token are mandatory
 
  \abstract
  % context heading (optional)
  % {} leave it empty if necessary  
   {Understanding dark matter halo dynamics can be pivotal in unraveling the nature of dark matter particles. Analytical treatment of the multistream flows inside the turn-around region of a collapsed CDM (cold dark matter) halo using various self-similar approaches already exist. In this work, we aim to determine the extent of self-similarity in two-dimensional halo dynamics and the factors leading to deviations from it by studying numerical simulations of monolithically growing CDM halos. We have adapted the Fillmore and Goldreich self-similar solutions assuming cylindrical symmetry to data from 2D Vlasov-Poisson (ColDICE package) simulations of CDM halos seeded by sine wave initial conditions. We measured trajectories in position and phase-space, mass and density profiles and compared these to predictions from the self-similar model, characterized by 2 parameters: $M_0, \epsilon$. The former scales the size of the turn-around region and the latter is inversely related to the mass accretion rate. We find that after turn-around and subsequent shell crossing, particles undergo a period of relaxation, typically about 1-2 oscillations about the center before they start to trace the self-similar fits and continue to do so as long as their orbits are predominantly radial. Overplotting the trajectories from different snapshots in scale-free position-time and phase spaces shows strikingly good superposition, a defining feature of self-similarity. The radial density profiles measured from simulations: $\rho \propto r^{-\alpha}, \alpha = 0.9 - 1.0$ are consistent with Fillmore and Goldreich's prediction $\alpha=1$ for 2D halos. The best-fit parameters for each simulation are found to be narrowly distributed, with the spread being entirely systematic. Deviations from self-similarity, on the other hand, are evidently linked to relaxation, inhibited motion due to periodic boundaries, transverse motion in the halo interior, and deficit of infalling mass in limited simulation volume. It could not be conclusively established if the halos tend to grow circular over time. Extension of this work to actual 3-dimensional CDM cosmologies necessitates further detailed study of self-similar solutions with elliptical collapse and transverse motion.}
  % aims heading (mandatory)
  % {}
  % methods heading (mandatory)
  % {}
  % results heading (mandatory)
  % {}
  % conclusions heading (optional), leave it empty if necessary 
  % {}


   \maketitle
%
%-------------------------------------------------------------------

\section{Introduction}
\label{sec:intro}

In the concordant model of cosmology, most of the mass of the universe $\sim$ 84\% \citep[see, e.g.][and references therein]{Planck_2020} is attributed to Cold Dark Matter (CDM), a non-relativistic collisionless fluid with negligible initial velocity dispersion that exhibits gravitational interactions. The large-scale structure of the universe is therefore dominated by the clustering of dark matter, with collapsed halos serving as the basic units of cosmological structures. The potential wells formed by these halos aid in the clustering of baryons, making them the birthing ground for stars and galaxies. The halo profile, crucial for modeling observables such as the annihilation signal of WIMPs \citep[e.g.,][]{Slatyer2022}, is very sensitive to the dynamics and the nature of dark matter particles. Thus, the study of dark matter halos and associated dynamics has important implications for dark matter indirect detection, the nature of dark matter particles, cosmological structure as well as galaxy formation and evolution.

The complex and multifaceted dynamics of dark matter particles is challenging to capture into a single-encompassing theory. Several analytical theories, each with its own set of assumptions and limitations, have been suggested for each stage of dynamics, allowing us to piece together a sketch of CDM structure formation in our universe. It begins with density perturbations set at the end of inflation that grow linearly, with particle motion dominated by Hubble expansion. After reaching a critical density, the particles decouple from the expansion, turn around and start collapsing. The single stream flow until the first shell-crossing (the intersection of particle trajectories) can be accurately captured by the Eulerian or Lagrangian perturbation theory \citep[see, e.g.,][and references therein]{Bernardeau_2002_review}. For simplified initial conditions composed of three crossed sine waves, Lagrangian perturbation theory (LPT) has also been shown to predict the crossing time and structure of phase-space and caustics (the loci of points where particle trajectories cross) \citep[see, e.g.,][]{Saga_2022}. After shell-crossing, the flow inside the splashback radius (the radius of the outermost caustic) turns multistream and LPT loses its validity. Nevertheless, the motion up to the next crossing time can still be followed by a post-collapse perturbative approach \citep{Colombi2015,TC_2017_1DPCPT, Rampf2021,STC_2023_3DPCPT}. Subsequent shell-crossings along transverse directions and violent relaxation in the multistream regime \citep{Lynden_1967} lead to the formation of primordial halos. In $N$-body simulations, these primordial halos are found to have a cusp-like density profile at the center obeying the approximate power law: $\rho \propto r^{-1.5}$ \citep[e.g.,][]{Diemand_2005,Ishiyama2010,Angulo2017,Colombi_2021,Delos_white_2023}. Accretion and mergers further drive the evolution of these halos towards a dynamical attractor, the well-known universal NFW profile \citep{NFW_1996,NFW_1997} or its recent improvements \citep[e.g., Einasto profile, see][]{Einasto_1965,Navarro_2004}.

Despite multiple attempts, there exists no complete analytical theory capable of fully describing the multistream dynamics and accurately predicting the slope and evolution of the density profile of CDM halos. One still promising class of such analytical approaches is to invoke self-similarity \citep[e.g.][but this list is not exhaustive]{Fillmore_1984,Bertschinger_1985,White_1992,Henriksen_1995,Sikivie_1997,Nusser_2001,Zukin_2010,Zukin_2010b,Lithwick_2011,Alard_2013,Sugiura_2020}. For a monolithically growing self-gravitating collisionless CDM halo in a matter dominated $\Omega_m = 1$ universe, the only characteristic scale is that set at turn-around, i.e. the instant of decoupling from cosmological expansion. Therefore, if position, time and mass profile were to be scaled with respect to the turn-around radius, time and mass inside the turn-around region, all the particles would trace the same trajectory regardless of their initial conditions i.e. the system would be 'self-similar'. \cite{Fillmore_1984}, hereby FG, developed self-similar solutions for planar, cylindrical and spherical symmetries. They were able to compute position-time and phase-space trajectories as well as mass and density profiles. For the cylindrically symmetric (2D) case, which will be the focus of this work, the predicted asymptotic density profile is $\rho \propto r^{-1}$, independent of the model parameters and for the spherically symmetric (3D) case, it varies between $\rho \propto r^{-2.25}$ - $r^{-2}$ depending on the halo mass. \cite{Bertschinger_1985} independently developed a self-similar theory of secondary spherical infall around an already collapsed density perturbation, which predicts a profile $\rho \propto r^{-2.25}$. There lies a clear contention between the density profiles predicted by these self-similar theories and those of prompt cusps and universal NFW profiles measured in CDM simulations. Many extensions to FG exist, that include, e.g., angular momentum \citep[e.g.,][]{White_1992,Sikivie_1997,Nusser_2001}, tidal torque \citep[e.g.,][]{Zukin_2010,Zukin_2010b} and ellipsoid dynamics \citep[]{Lithwick_2011}. In this paper, we focus on the FG self-similar model, aware of the inherent limitations due to the assumed symmetry and purely radial motion.

CDM is modelled as a self-gravitating and collisionless fluid obeying Vlasov-Poisson equations:
\begin{align}
    &\frac{\partial f}{\partial t} + \mathbf{u}\cdot \nabla_{\mathbf{r}}f - \nabla_{\mathbf{r}} \phi \cdot \nabla_{\mathbf{u}}f = 0,\\
    &\Delta_{\mathbf{r}}\phi = 4\pi G\, \rho = 4\pi G \int f(\mathbf{r}, \mathbf{u}, t)\, {\rm d}^3\mathbf{u},
\end{align}
where $f(\mathbf{r}, \mathbf{u}, t)$ represents the phase-space density at position $\mathbf{r}$, velocity $\mathbf{u}$ and time $t$ and $\phi$ is the gravitational potential. These are traditionally resolved with an $N$-body approach i.e. representing the phase-space by an ensemble of particles which follow the standard Lagrangian equations of motion in an expanding universe:
\begin{equation}
    \label{eq:lagrange_eq_motion} \frac{{\rm d}^2\mathbf{x}}{{\rm d}t^2} + 2H\frac{{\rm d}\mathbf{x}}{{\rm d}t} = -\frac{1}{a^2} \mathbf{\nabla}_{\mathbf{x}}\phi(\mathbf{x}) \hspace{0.5 cm} ; \hspace{0.2 cm} \Delta_{\mathbf{x}} \phi(\mathbf{x}) = 4\pi G \,\bar{\rho}\, a^2\, \delta(\mathbf{x}),
\end{equation}
where $\mathbf{x}$ is the comoving position, $a$ is the cosmological expansion factor corresponding to time $t$ and $\delta$ is the matter density contrast, defined as $\delta = \rho/\bar{\rho} - 1$ with $\bar{\rho}$ being the background matter density.

In this work, we opt to study dark matter halo dynamics in 2D Vlasov simulations of simplified initial conditions composed of two crossed sine waves run using the ColDICE Vlasov solver of \citet{Sousbie_2016}. These highly symmetrical setups with predetermined halo centers, no mergers or angular momentum are ideal for testing simplistic self-similar models like that of FG. For comparing to our 2D simulations, we use the cylindrically symmetric FG solutions which are known to lead to a parameter-independent asymptotic profile $\rho \propto r^{-1}$ that can be easily verified with our simulations. Though the 2D dynamics are much simpler than in an actual 3D CDM cosmology, it makes the study of fine deviations from self-similar behaviour and their causes much more feasible. FG self-similar solutions have already been used to compare against phase-space structures \citep{Sugiura_2020} and density profiles \citep{Enomoto_2023_letter, Enomoto_2023_full} of late-time equilibrated CDM halos in the multistream regime. Our focus shall be on probing deeper at the level of individual particle trajectories starting from the first shell-crossing itself in addition to investigating phase-space, transverse motion and profiles of mass, density and anisotropy parameter as measured in our simulations. As an additional note, dynamics in 2D corresponds to the interaction between infinite lines in 3D, which can be viewed as an approximation of the dynamics of filaments in the context of 3D CDM cosmological dynamics.

The plan of the paper is as follows. In section \ref{sec:numerical_sim}, we detail the methodology used to conduct numerical simulations and generate data for isolated halos. In section \ref{sec:theory}, we revisit the FG self-similar model and show the semi-analytic solutions for trajectories, mass and density profiles. In section \ref{sec:data_comp}, we outline our choice of observables and the procedure of fitting self-similar solutions to our numerical data to obtain bounds and best-fit parameters, additionally discussing the causes behind deviations from self-similarity. The resulting distribution of best-fit parameters and their comparision across simulations with different initial set-ups are discussed in section \ref{sec:param_dist}. A summary of our key qualitative inferences about dark matter halo dynamics in 2D, possible theoretical extensions and its implications on actual 3D CDM cosmological simulations and observations are presented in section \ref{sec:conclusions}.
%--------------------------------------------------------------------
\section{Numerical simulations}
\label{sec:numerical_sim}

\begin{figure*}
   \centering
   \includegraphics[width = 0.33\textwidth]{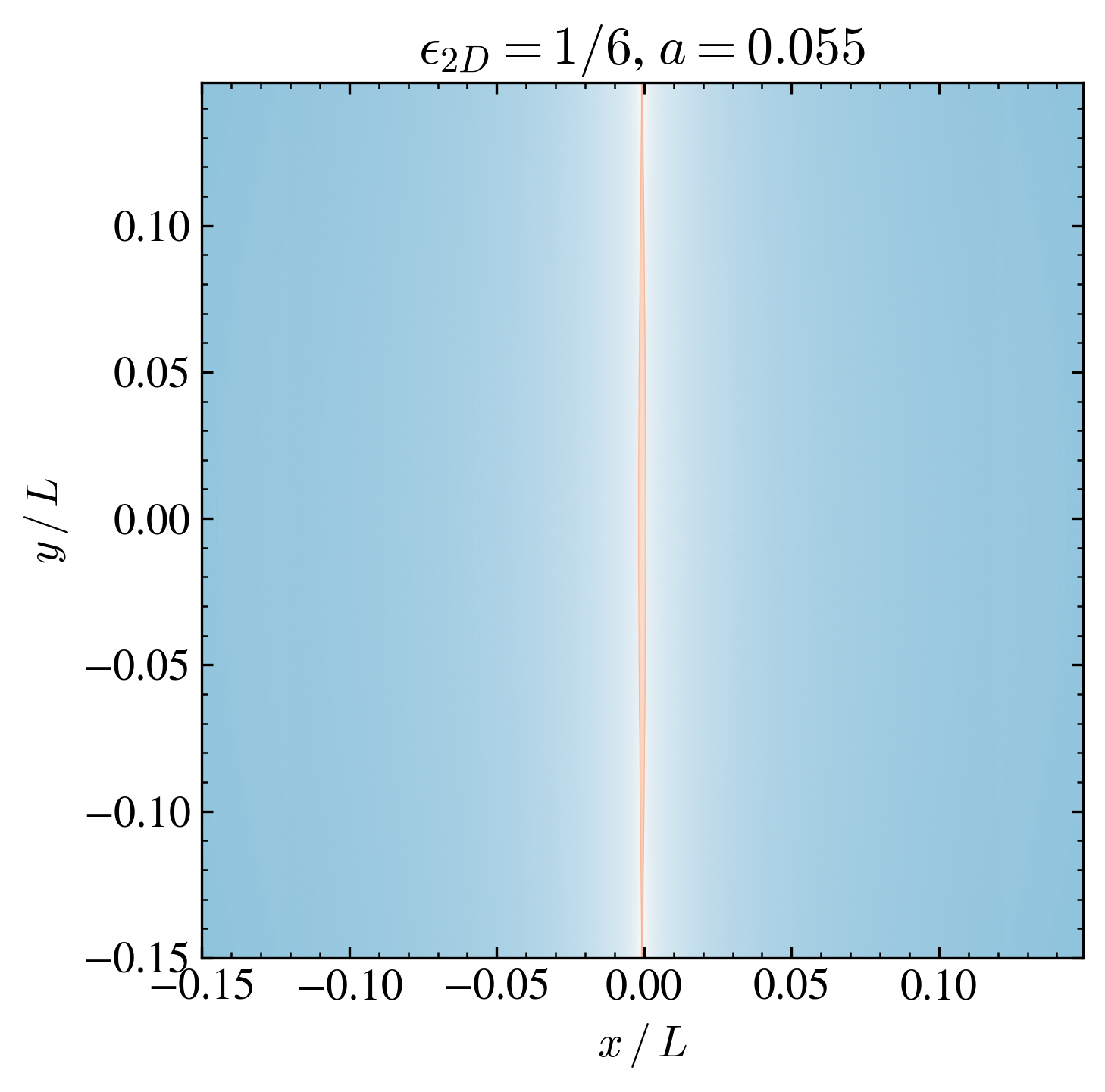}
   \includegraphics[width = 0.33\textwidth]{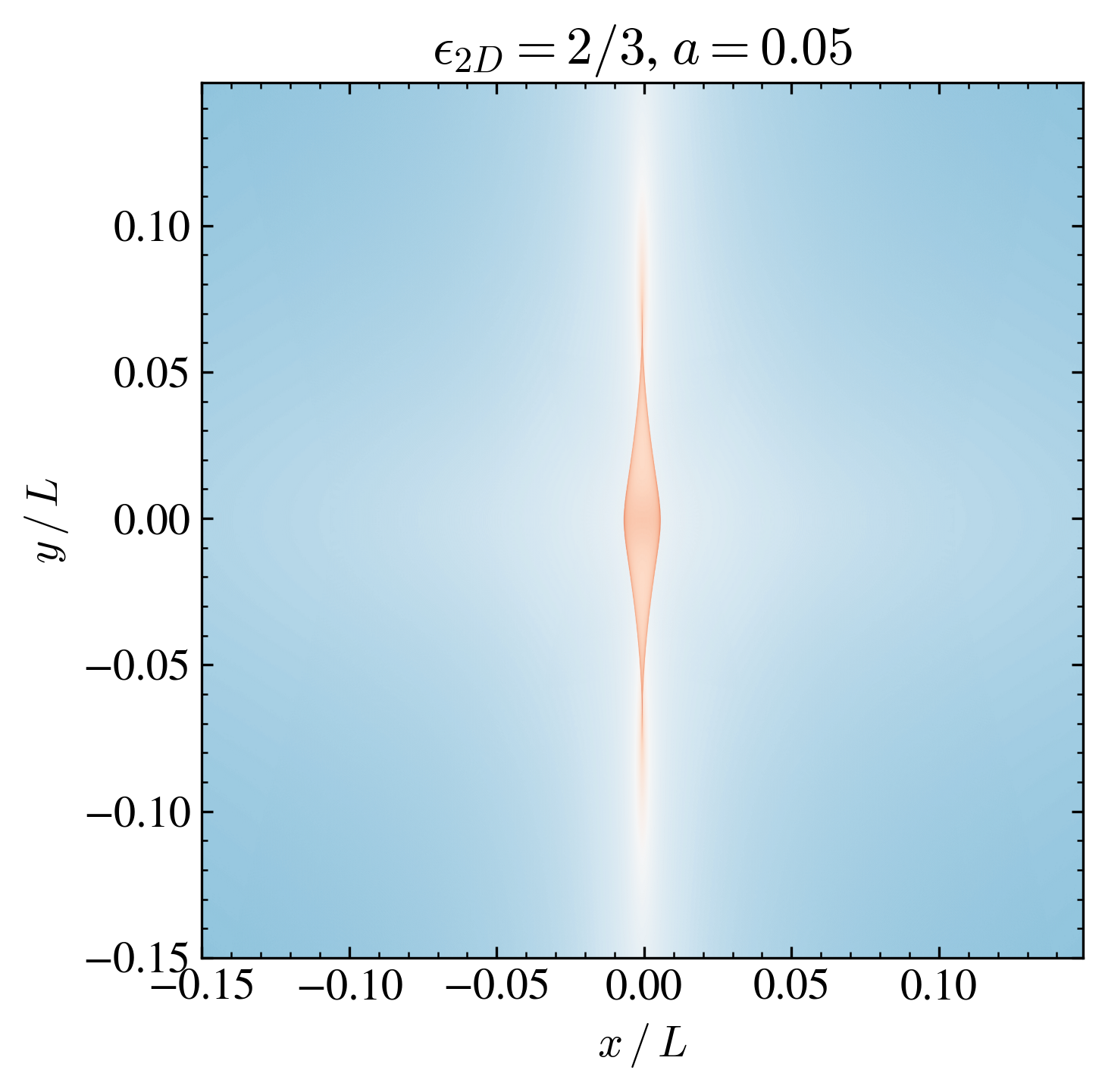}
   \includegraphics[width = 0.33\textwidth]{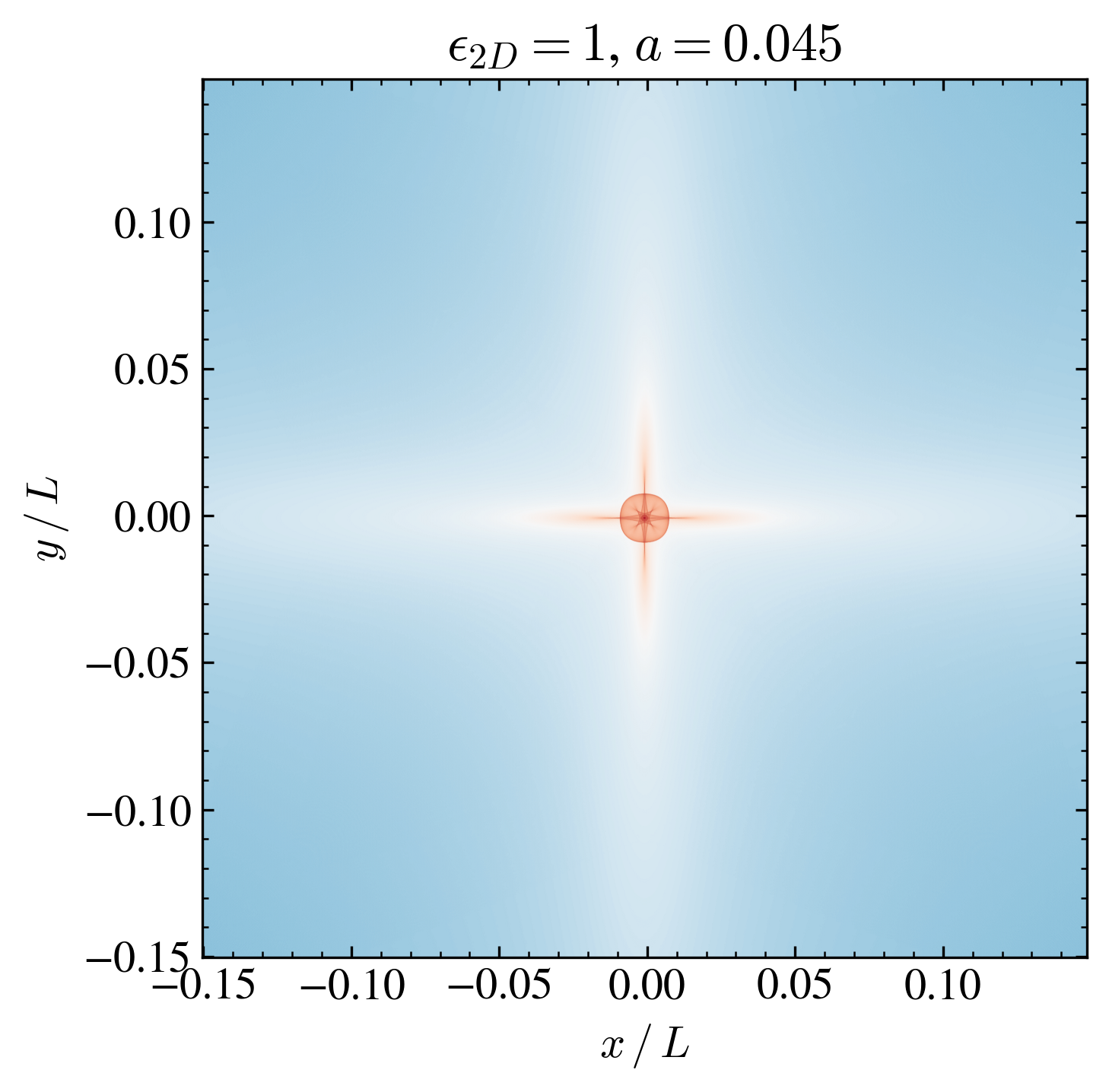}

   \includegraphics[width = 0.33\textwidth]{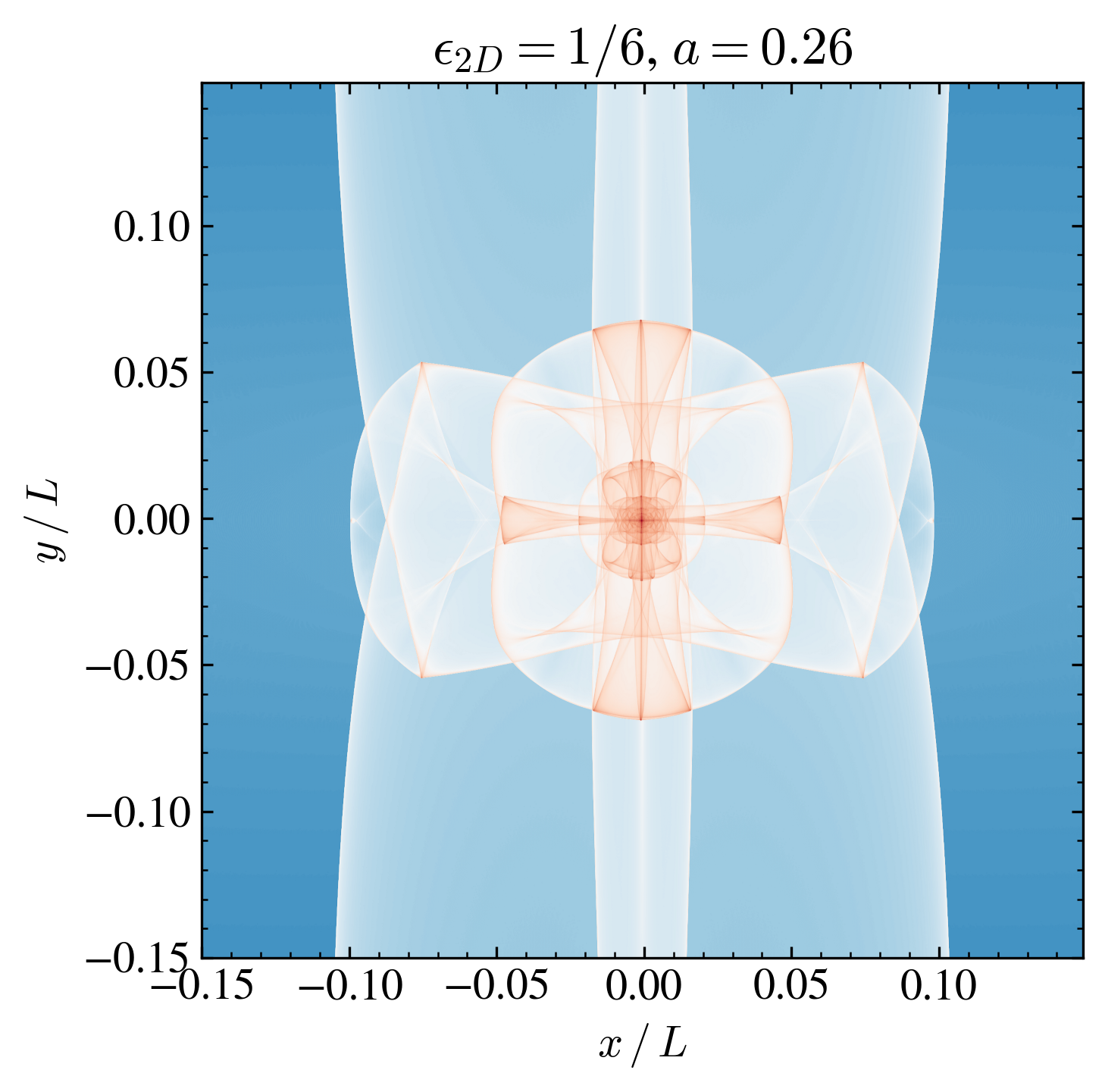}
   \includegraphics[width = 0.33\textwidth]{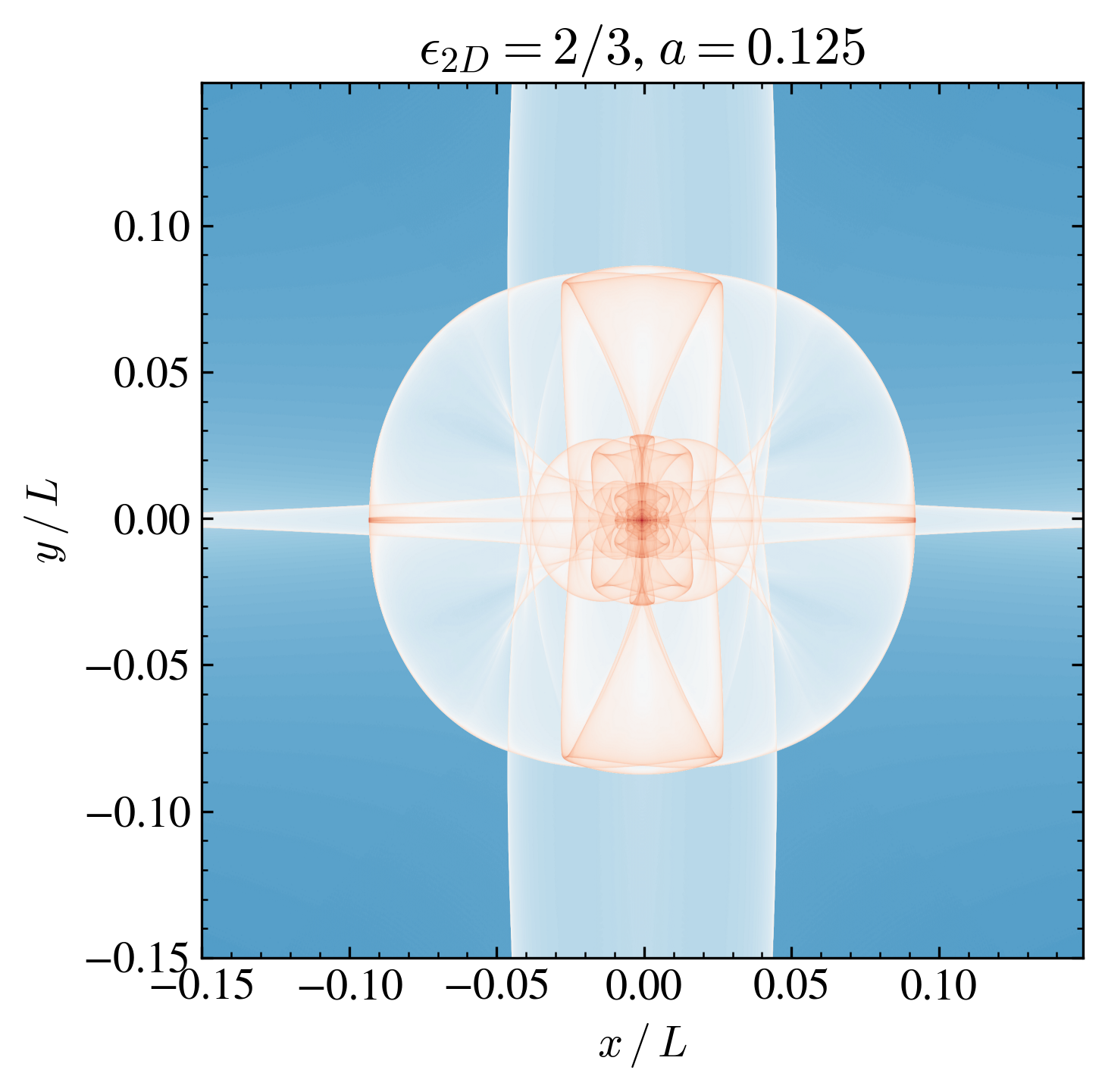}
   \includegraphics[width = 0.33\textwidth]{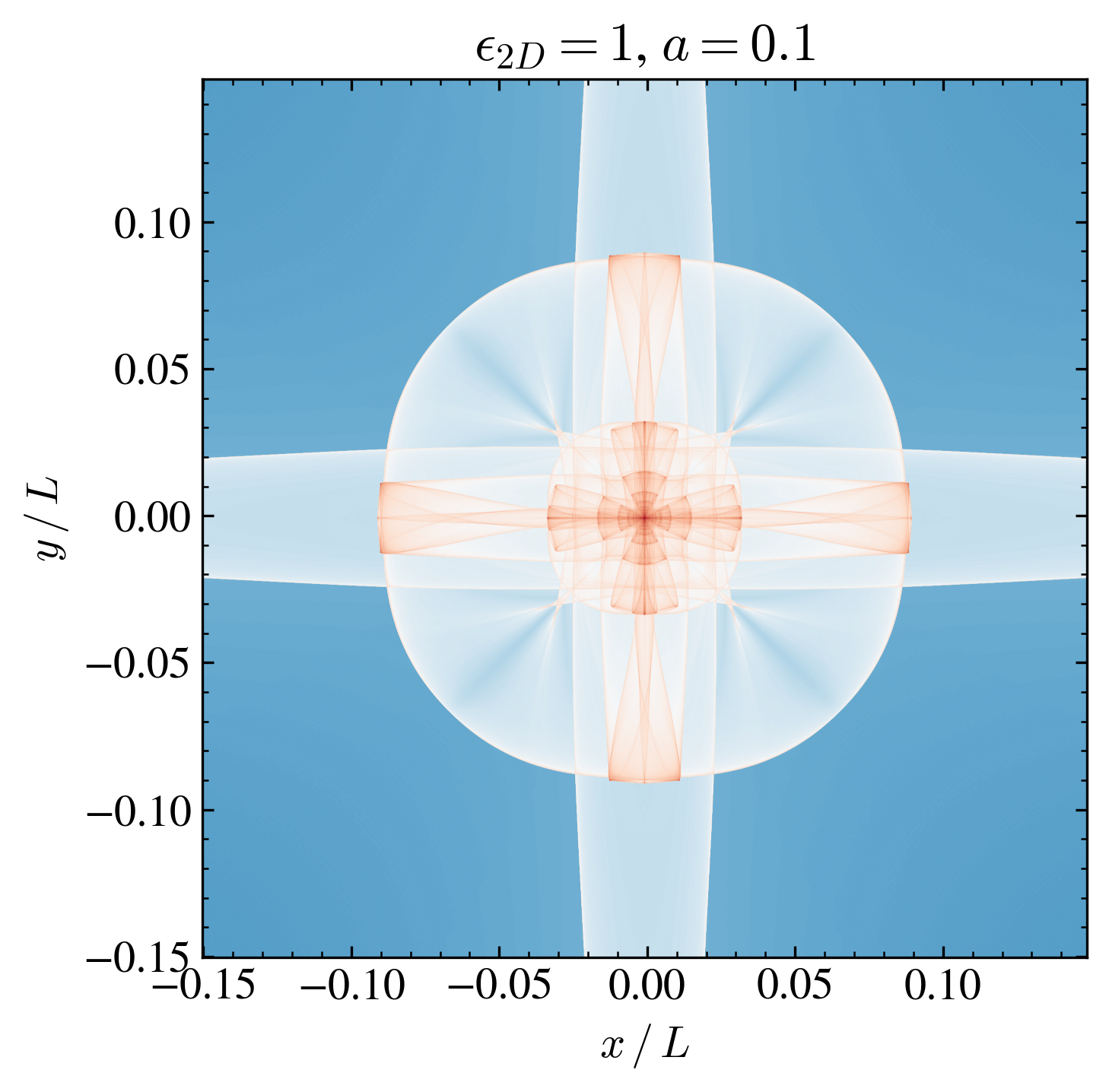}

   \includegraphics[width = 0.33\textwidth]{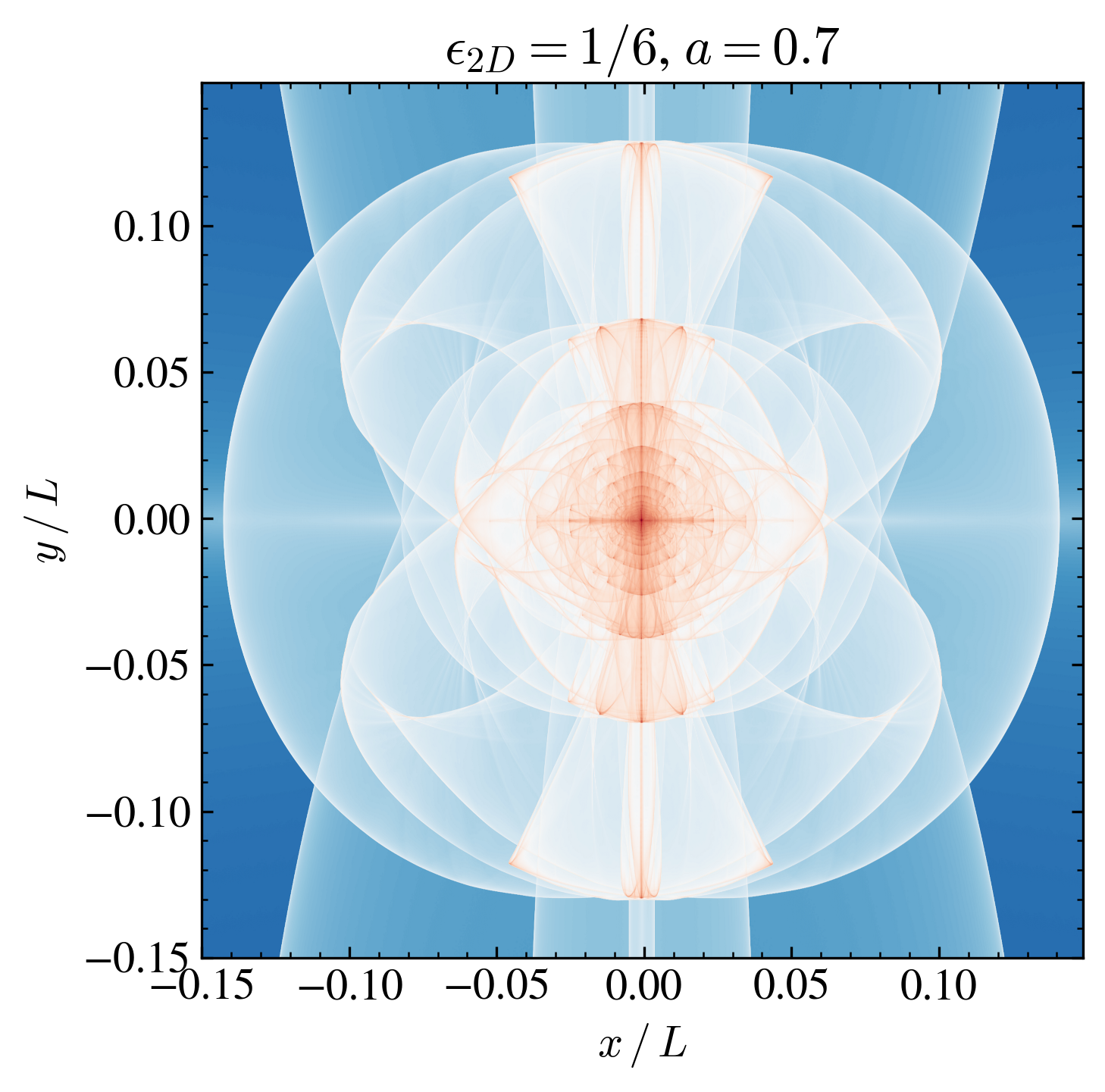}
   \includegraphics[width = 0.33\textwidth]{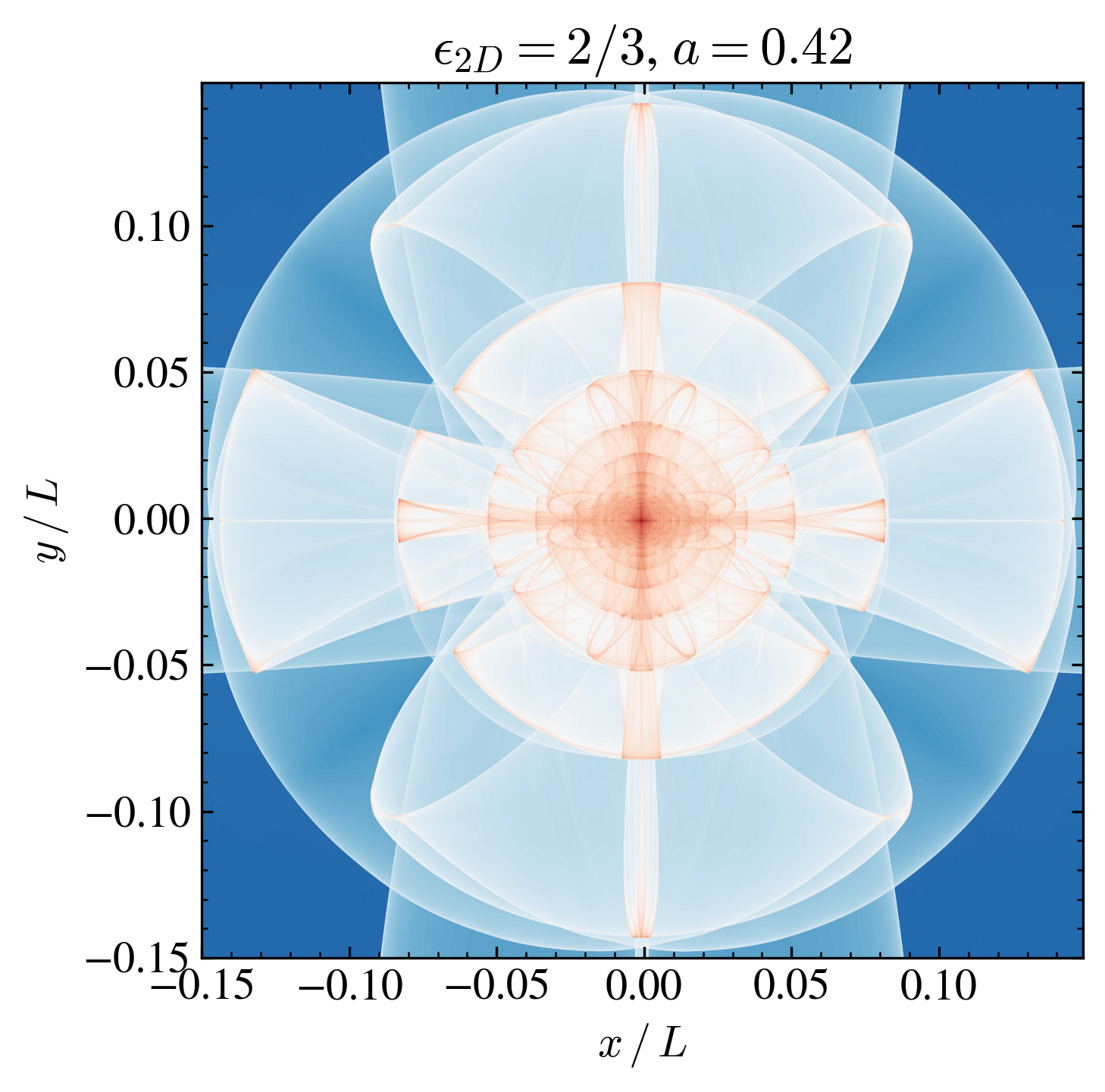}
   \includegraphics[width = 0.33\textwidth]{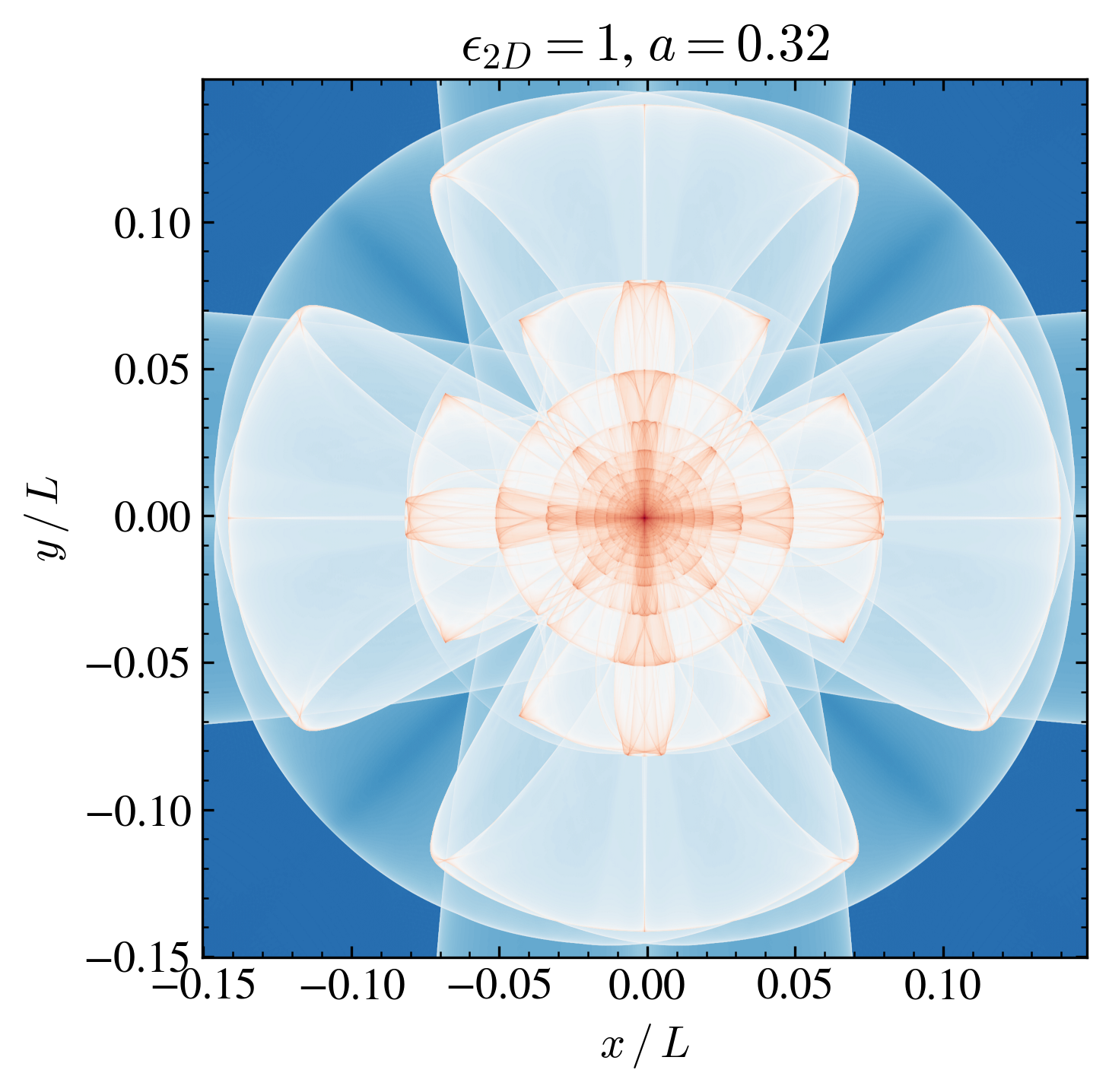}
   \caption{Density colormaps of our numerically simulated halos. In the left column, we have Q1D simulation snapshots for the expansion factors $a = 0.055, 0.26, 0.7$. In the middle, we have ANI simulation snapshots for $a = 0.05, 0.125, 0.42$. In the right column, we have SYM simulation snapshots for $a = 0.045, 0.1, 0.32$. The top row consists of the closest available snapshots after the first shell crossing in each of the simulations and the bottom row consists of the last available snapshots.}
   \label{fig:2SIN_vl_snaps}
\end{figure*}

The simulations under study were carried out by \cite{Saga_2022} using the public Vlasov solver ColDICE \citep{Sousbie_2016}. It models the CDM phase-space distribution $f$ as a 2D (or 3D) sheet, assuming negligible initial velocity dispersion, in a 4D (or 6D) phase-space:
\begin{align}
    f(\mathbf{r}, \mathbf{u}, t = t_{\rm ini}) = \rho_{\rm ini}(\mathbf{r})\, \delta_{\rm D}(\mathbf{u} - \mathbf{u_{\rm ini}}),
\end{align}
where $\rho_{\rm ini}, \mathbf{u_{\rm ini}}$ are the density and peculiar velocity fields initialised by Zeldovich flow. The phase-space sheet is adaptively tessellated with simplices (triangles in 2D and tetrahedra in 3D) whose vertices $[\mathbf{x}(t), \mathbf{v}(t)]$ are then evolved according to the Lagrangian equations of motion \eqref{eq:lagrange_eq_motion}. The matter is linearly distributed inside each simplex instead of being transported by the vertices, unlike the $N$-body approach. For more details on refinement and measurements, refer to \cite{Sousbie_2016} and \cite{Saga_2022}.

We choose to study highly symmetric cases, where the displacement field $\mathbf{\Psi}$ is initialised by sine waves with amplitudes ($\epsilon_x, \epsilon_y$) along $x$ and $y$ axes:
\begin{align}
    &\mathbf{x}(\mathbf{q}, t) = \mathbf{q} + \mathbf{\Psi}(\mathbf{q}, t),\\
    &\Psi_{\rm i}(\mathbf{q}, t_{\rm ini}) = \frac{L}{2\pi} D_{+}(t_{\rm ini})\, \epsilon_{\rm i}\, \sin{\left( \frac{2\pi}{L} q_{\rm i} \right)},
\end{align}
where $\mathbf{q}$ is the comoving initial position, $D_+$ is the linear growth factor, $L$ is the comoving size of the simulation box with periodic boundaries and ${\rm i}$ is an index for $x, y$. The three sets of ($\epsilon_x, \epsilon_y$) = (-18, -3), (-18, -12), (-18, -18) are used to set the initial conditions for the simulations, which we denote as quasi-1D (Q1D), anisotropic (ANI) and axial-symmetric (SYM) respectively. The parameters used for each simulation are tabulated in Table \ref{table:vl_sim_params}.  The first shell-crossing (self-intersection of the phase-space sheet) occurs along the $x$-axis, followed by shell-crossing along the transverse $y$ direction that leads to the formation of a single monolithically growing halo precisely at the center, which is the ideal scenario to test for self-similar particle trajectories unaffected by mergers. The simulations are in increasing order of resemblance to circular symmetry which is assumed in FG self-similar solutions. Our comparative study across the three simulations allows us to investigate how the halo properties vary with initial conditions.   

\begin{table*}
    \centering
    \begin{tabular}{cccccccccc}
        \hline
        Designation & $\epsilon_{\rm 2D}$ & $n_{\rm g}$ & $n_{\rm s}$ & $a_{\rm i}$ & $a_{{\rm SC}, x}$ & $a_{{\rm SC}, y}$ & $a_{\mathcal{M}}$ & $a_{r}$ \\ \hline
        Q1D & 1/6   & 2048 & 2048 & 0.005 & 0.05285 & 0.14 & 0.2 & 0.26 \\
        ANI & 2/3   & 2048 & 2048 & 0.005 & 0.04545 & 0.055 & 0.1 & 0.125\\ 
        SYM & 1     & 2048 & 2048 & 0.005 & 0.04090 & 0.04090 & 0.055 & 0.075\\ \hline
    \end{tabular}
    \caption{ Parameters of the three simulations analyzed in this article. All three simulations have been performed assuming a matter-dominated universe $\Omega_{\rm M} = 1$ and starting expansion factor $a_{\rm ini} = 0.0005$. The second column gives the relative amplitude of the sine waves along $x$ and $y$ axes: $\epsilon_{\rm 2D} = \epsilon_y/\epsilon_x$. $n_{\rm g}$ denotes the spatial resolution of the grid over which the Poisson equation is solved and $n_{\rm s}$ denotes the spatial resolution of the grid used for tessellation. $a_{\rm i}$ is the expansion factor of the first output snapshot, whose mass profile $M_{\rm i}$ we measure and scale our self-similar solutions accordingly. $a_{{\rm SC}, x}$, $a_{{\rm SC}, y}$ are the estimated expansion factors at first shell-crossings along $x, y$ axes respectively. $a_{{\mathcal{M}}}$, $a_{r}$ are the expansion factors of the earliest snapshots for which self-similar fits to the mass profile and the particle trajectories could converge.}
    \label{table:vl_sim_params}
\end{table*}
Figure \ref{fig:2SIN_vl_snaps} shows the density projections of the halos seeded by the different initial conditions in each simulation at gradually increasing times. The structure of caustics i.e. the points corresponding to the folds in the phase-space sheet with singular density, is clearly visible. The most notable feature about their evolution is that despite the stark symmetry difference in the shape of the caustics at shell-crossing, they grow roughly similar and close to circular at late times.
    
%--------------------------------------------------------------------
\section{Theory}
\label{sec:theory}

\begin{figure*}
    \includegraphics[width = \textwidth]{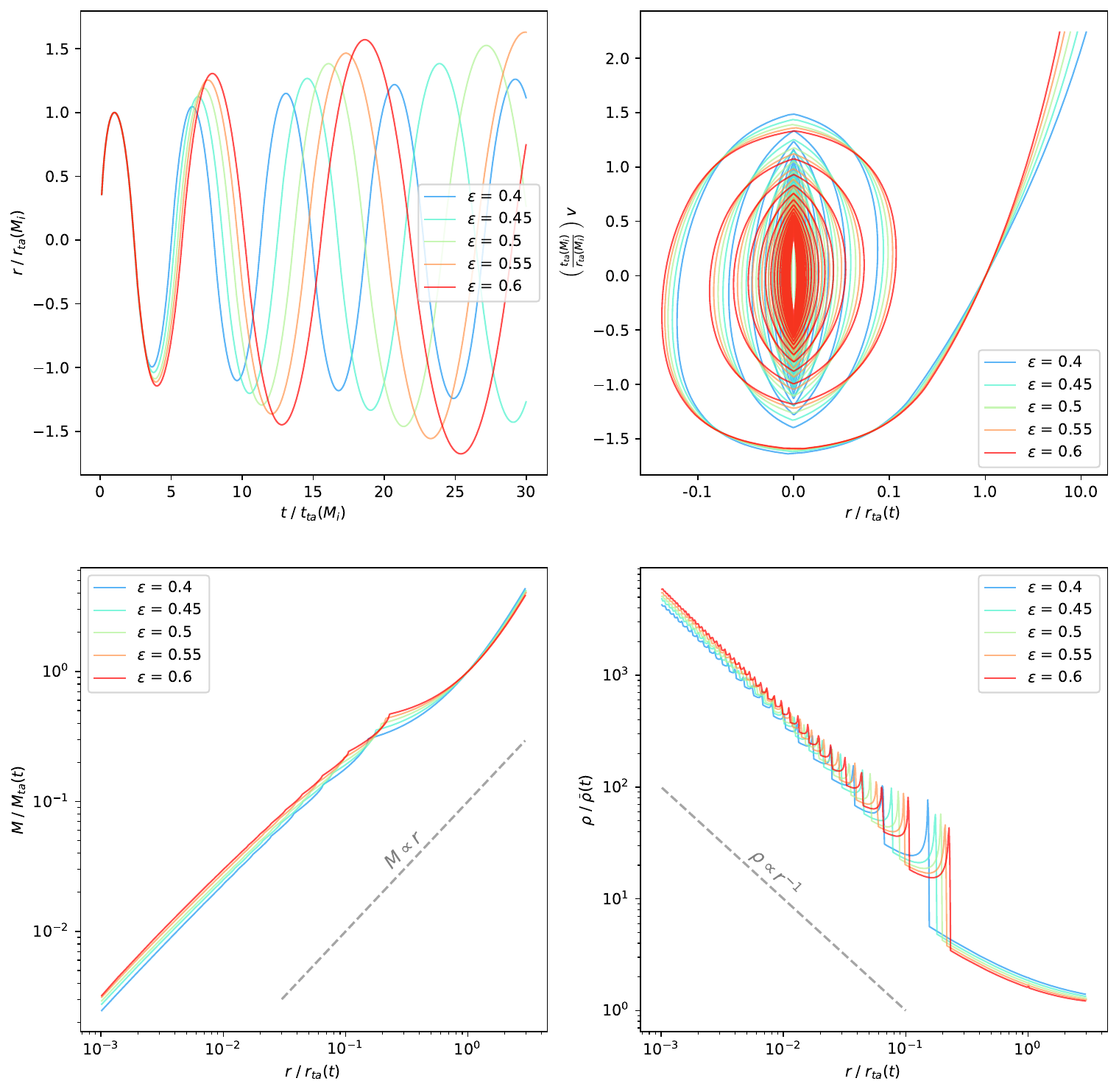}
    \centering
    \caption{FG self-similar solutions assuming cylindrical symmetry ($n = 2$) for $\epsilon \in [0.4, 0.6]$. The solutions are depicted in scale-free spaces(variables normalised w.r.t the turnaround). {\it Top left:} position - time trajectories. {\it Top right:} phase-space trajectories. {\it Bottom left:} mass profiles. {\it Bottom right:} density profiles.}
    \label{fig:self_sim_sol}
  \end{figure*}

We briefly recap the FG self-similar solutions. A starting expansion factor $a_{\rm i}$ in the matter domination era wherein the flow is dominated by Hubble expansion and a scale-free (power-law) initial perturbation $\delta$ are assumed:

\begin{equation}
    \centering
    \label{eq:init_delta_ss} \delta \equiv \frac{\delta M_{\rm i}}{ M_{\rm i}} = \left( \frac{M_{\rm i}}{M_0} \right)^{-\epsilon} \hspace{0.5 cm};\hspace{0.2 cm} M_{\rm i} \equiv M(r, t_{\rm i}) = \int_0^r r'^{n-1} \rho(r', t_{\rm i}) \; {\rm d}r',
\end{equation}
\noindent where $M_{\rm i}$ is the initial profile of cumulative mass in planar $(n = 1)$, cylindrical $(n = 2)$ or spherical $(n = 3)$ shells and $M_0, \epsilon$ are the model parameters. $\epsilon$ is related to the halo mass and accretion rate and $M_0$ is a reference mass. We select the units $\bar{\rho}(t_{\rm i}) (a_{\rm i} L)^n$ for $M_{\rm i}, M_0$ so that their values are independent of the length and mass scales used and physically represent the fraction of the total mass. For the first output snapshots of our simulations with $a_{\rm i} = 0.005$, the assumption of the initial motion being dominated by the Hubble flow remains valid, but not that of the initial perturbation $\delta$ obeying a power law. However, it is a plausible assumption that the trajectories with small $\delta$ do approach a self-similar solution gradually.

Thus, we start in the single-stream regime, where the particles (analogous to points in the phase-space sheet in the Vlasov simulations) continue to expand away from the center. Once the local gravitational pull dominates over the Hubble expansion, the particles turn around. Since the average density decreases with distance from the center, particles initially farther away turn around later i.e. the turnaround radius grows with time. Subsequently, particles undergo first shell-crossing symmetrically along all directions leading to the formation of the first caustics. The flow inside the splashback radius (radius of the outermost caustic) turns multi-stream. The particles in this regime continue to oscillate about the center with an asymptotically converging amplitude. As particles farther away from the center continue to turn around and infall, a power-law density profile builds up in the multistream regime.

As discussed in section \ref{sec:intro}, the system possesses only one characteristic scale, which is defined by the turnaround. The variables in the system - position $r$ and time elapsed $t$ for particles (labelled by the initial mass $M_{\rm i}$ enclosed in concentric shells on which they are located) and mass profile $M(r, t)$ - are normalised w.r.t. the values at turnaround:

\begin{align}
    \label{eq:tau}&\tau = \frac{t}{t_{\rm ta}(M_{\rm i})} = C_t^{-3/2} \left( \frac{t}{t_{\rm i}} \right) \left( \frac{M_0}{M_{\rm i}} \right)^{3\epsilon/2},\\
    \label{eq:lambda}&\lambda = \frac{r}{r_{\rm ta}(M_{\rm i})} = C_r^{-1} \left( \frac{r}{r_{\rm i}} \right)\left( \frac{M_0}{M_{\rm i}} \right)^{\epsilon},\\
    \label{eq:M}&\mathcal{M}\left( \frac{r}{r_{\rm ta}(t)} \right) = \frac{M(r, t)}{M_{\rm ta}(t)} \hspace{0.5 cm} ; \hspace{0.1 cm} M_{\rm ta}(t) = \frac{M_0}{C_t^{1/\epsilon}} \left( \frac{t}{t_{\rm i}} \right)^{2/3\epsilon - 2(3-n)/3},
\end{align}
\noindent where the indices $\rm ta$ and $\rm i$ refer to turnaround and initial. Our case of interest is that of cylindrical symmetry $n = 2$, for which the coefficients $C_r$, $C_t$ relating turnaround radius $r_{\rm ta}$ and time $t_{\rm ta}$ to the initial density perturbation $\delta$ are $0.74$ and $1.39$, respectively. It is important to recognize that $r_{\rm ta}(M_{\rm i})$, the turnaround radius of the shell enclosing initial mass $M_{\rm i}$, is different from $r_{\rm ta}(t)$ which is the radius of shell turning around at the instant $t$. The coupled differential equations in terms of the scaled variables are as follows:
\begin{align}
    \label{eq:main_eq_lambda}&\frac{{\rm d}^2 \lambda}{{\rm d}\tau^2} = \frac{2(3-n)}{9n} \frac{\lambda}{\tau^2} - \frac{2}{3n} \left( \frac{C_t}{C_r} \right)^n \tau^{2/3\epsilon - 2(3-n)/3} \frac{\lambda}{|\lambda|^n} \mathcal{M} \left( \frac{\lambda}{\Lambda} \right),\\
    \label{eq:main_eq_mass}&\mathcal{M} \left( \frac{\lambda(\tau)}{\Lambda(\tau)} \right) = \frac{2}{3 \epsilon} \int_1^{\infty} \frac{{\rm d}\tau'}{\tau'^{1+2/3\epsilon}} H\left[ \left| \frac{\lambda(\tau)}{\Lambda(\tau)} \right| - \left| \frac{\lambda(\tau')}{\Lambda(\tau')} \right| \right],
\end{align}
\noindent
where $\Lambda(\tau) = \tau^{2(1 + 1/3\epsilon)/3}$ and $H$ is the Heaviside step function. The boundary conditions are $\lambda(\tau = 1) = 1$ and $d\lambda / d\tau (\tau = 1) = 0$ . Note: the equations have been slightly modified by taking the absolute magnitude of $\lambda$ to change the sign of the particle's position upon crossing through the center $\lambda = 0$. This allows for smooth integration of the equations about the center.

The fact that the eqs. \eqref{eq:main_eq_lambda} and \eqref{eq:main_eq_mass} as well as the boundary conditions are independent of initial time $t_{\rm i}$, position $r_{\rm i}$ or enclosed mass $M_{\rm i}$ implies that all the particle trajectories trace the same curve in the scale-free space i.e. they are 'self-similar'. We need to solve the equations only once and then appropriately scale the solution to obtain the trajectory of any particle in our simulation.

Before turnaround ($\tau < 1$), we integrate backwards over a grid of $\tau \in [0, 1]$ with $d\tau = 0.01$. We substitute $\mathcal{M} = \tau^{-2/3\epsilon}$ since the initial mass enclosed remains conserved $M(r(t), t) = M_{\rm i}$ for single stream flow. After turnaround ($\tau \ge 1$), the equations are solved iteratively up to the desired convergence starting with an initial guess for $\mathcal{M}(\lambda/\Lambda)$ which is a monotonic function between $\mathcal{M}(0) = 0$ and $\mathcal{M}(1) = 1$. Eq. \eqref{eq:main_eq_lambda} is integrated forwards over a grid of $\tau \in [1, \tau_{\rm f}]$ with $d\tau = 0.01$ and the solution $\lambda(\tau)$ is linearly interpolated. $\tau_{\rm f}$ is chosen such that the amplitude of oscillations decreases below $\lambda/\Lambda = 10^{-3}$, allowing us to precisely resolve the mass profile down to $r/r_{\rm ta} = 10^{-3}$. The mass profile $\mathcal{M}$ is computed over a grid of 500 points $\in [10^{-3}, 1]$ and extrapolated down to $\mathcal{M}(0) = 0$ before being used for the next iteration.

The equation for computing the mass profile \eqref{eq:main_eq_mass} can be further simplified by breaking the integral into intervals of $\tau_k$'s which satisfy $\lambda(\tau)/\Lambda(\tau) = \lambda(\tau_k)/\Lambda(\tau_k)$ \citep{Bertschinger_1985}:
\begin{equation}
    \label{eq:main_eq_mass_simp}\mathcal{M}(\lambda/\Lambda) = \Sigma_{k = 1}^N (-1)^{k+1} \tau_k^{-2/3\epsilon}.
  \end{equation}
Numerically integrated solutions to the coupled eqs. \eqref{eq:main_eq_lambda} and \eqref{eq:main_eq_mass_simp} corresponding to the parameter $\epsilon = 0.4, 0.45, 0.5, 0.55, 0.6$ are presented in figure \ref{fig:self_sim_sol}. The parameter $M_0$ enters only while scaling the self-similar solutions back to the appropriate particle. From the plots of self-similar trajectories, we note that lower $\epsilon$ corresponds to lower amplitude and greater frequency of oscillations as well as lower velocity at center crossing. A higher frequency of oscillations implies a larger number of caustics, seen as spikes in mass and density profiles. The density profile is given by:

\begin{equation}
    \label{eq:density_profile}
   \frac{\rho(r, t)}{\bar{\rho}(t)} = \frac{1}{n} \left( \frac{C_t}{C_r} \right)^n \left( \frac{r_{\rm ta}}{r} \right)^{n-1} \frac{{\rm d} \mathcal{M} (r/r_{\rm ta}(t))}{{\rm d}(r/r_{\rm ta})}.
\end{equation}

\noindent The asymptotic logarithmic slope for the density profile is $-1$ for cylindrical symmetry $n = 2$ and is independent of $\epsilon$ (refer equation 39 of FG), which agrees with the numerical solution.

%--------------------------------------------------------------------
\section{Data Comparision}
\label{sec:data_comp}

In this section, we test the FG self-similar solutions assuming cylindrical symmetry against data from the 2D Vlasov simulations detailed in \ref{sec:numerical_sim}. We particularly focus on particle trajectories, mass and density profiles while also investigating the phase-space, transverse motion and anisotropy parameter. The key questions we want to address are the following:
\begin{enumerate}
    \item What is the extent of self-similarity in the motion of particles in our simulations?
    \item Are the mass and density profiles in agreement with self-similar predictions?
    \item Where do the deviations from self-similarity occur and what causes them?
    \item What is the distribution of best-fit parameters $M_0, \epsilon$ ? How does it compare across the three simulations? Do the halos become more circular with time?
    \item What are the implications on CDM halos seeded by Gaussian random fields in 3D cosmological simulations?
\end{enumerate}

\subsection{Particle trajectories}
\label{subsec:part_traj}

\begin{figure*}
    \includegraphics[width = 0.33\textwidth]{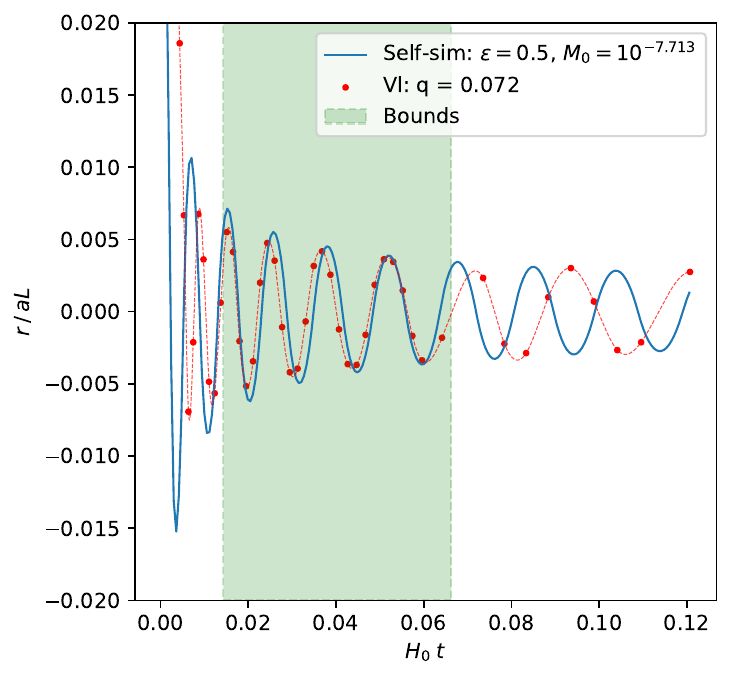}
    \includegraphics[width = 0.33\textwidth]{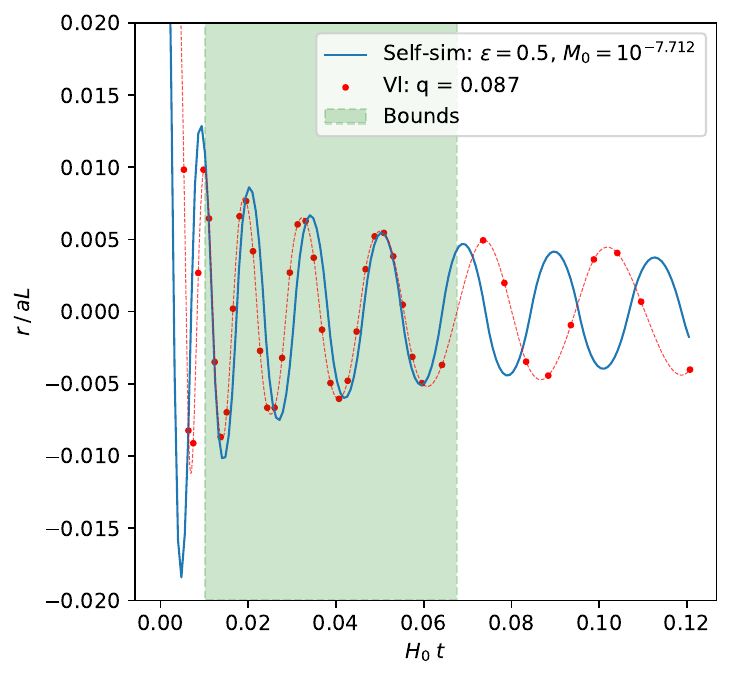}
    \includegraphics[width = 0.33\textwidth]{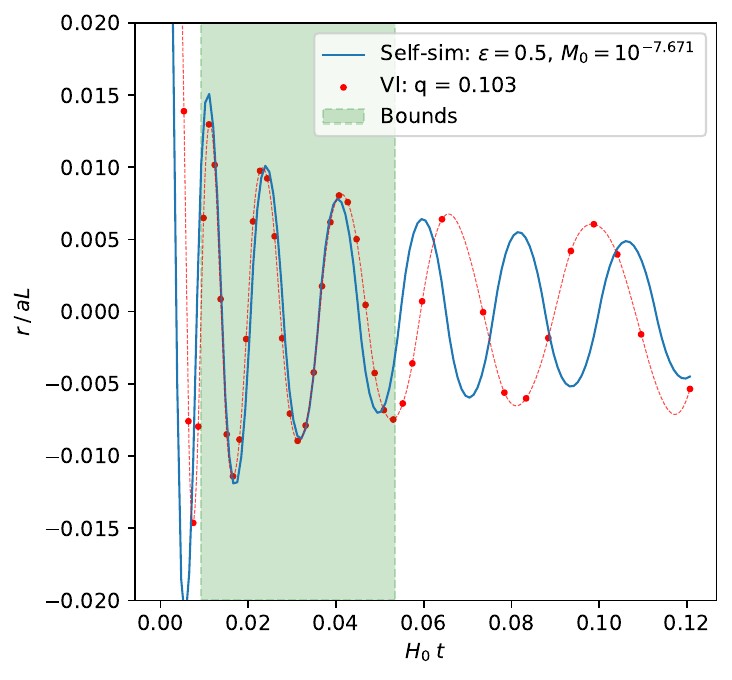}
    \centering
    \caption{Position-time trajectories of three Lagrangian points along $x$ axis: $q \,/\,L \: =$ 0.072, 0.087, 0.103 (left to right) in the SYM simulation. The data points (red, labelled 'Vl' as in Vlasov solver used for the simulations) have been cubically interpolated. The self-similar fits, made using the least-squares method with $\epsilon = 0.5$, are shown in blue. The time intervals (green) within which the trajectories follow the fits were computed assuming a threshold: $\Delta r/r_{\rm data} \le 10\%$. The spatial resolution of the tessellation grid is $\sim 0.0005$ comoving box units.}
    \label{fig:vl_comp_r_t}
\end{figure*}

\begin{figure*}
    \includegraphics[width = 0.33\textwidth]{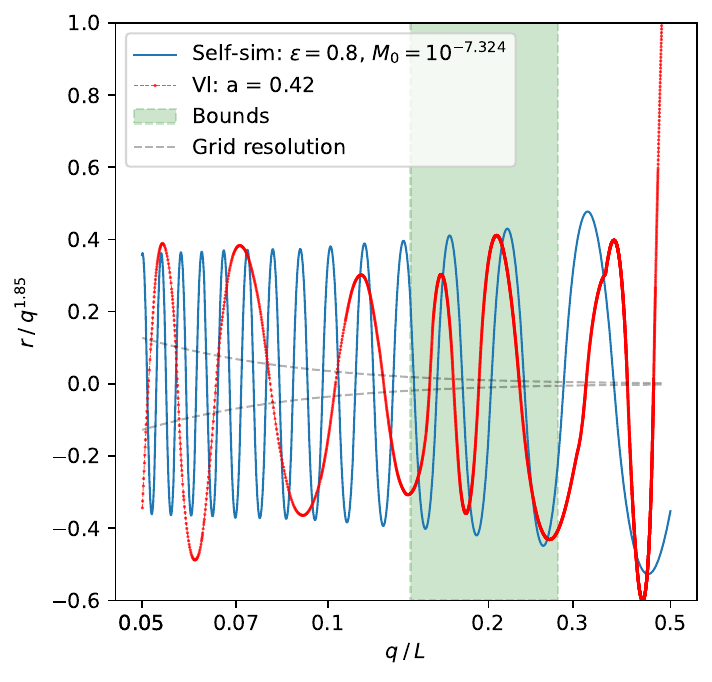}
    \includegraphics[width = 0.33\textwidth]{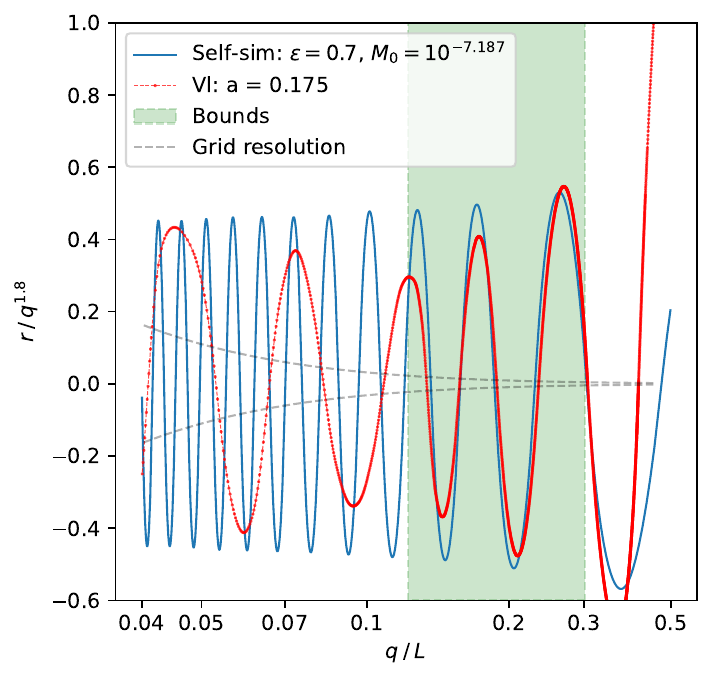}
    \includegraphics[width = 0.33\textwidth]{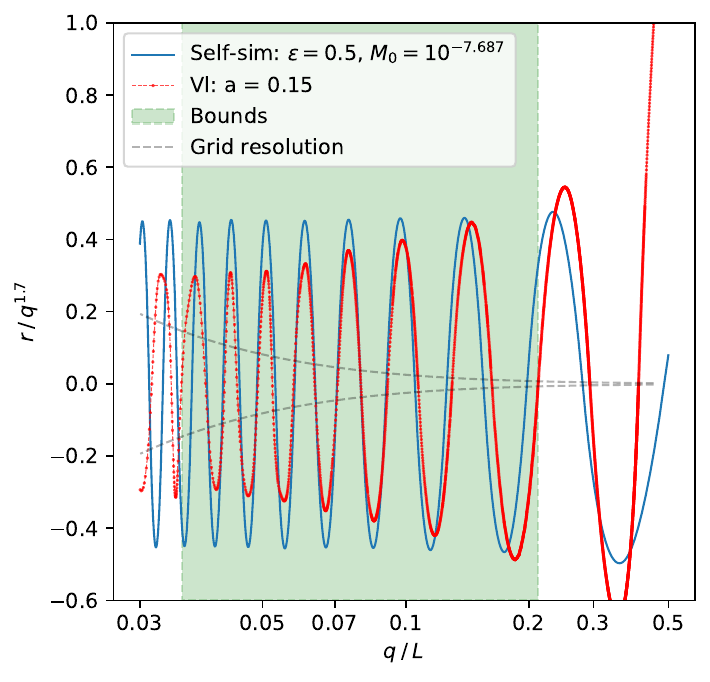}

    \includegraphics[width = 0.33\textwidth]{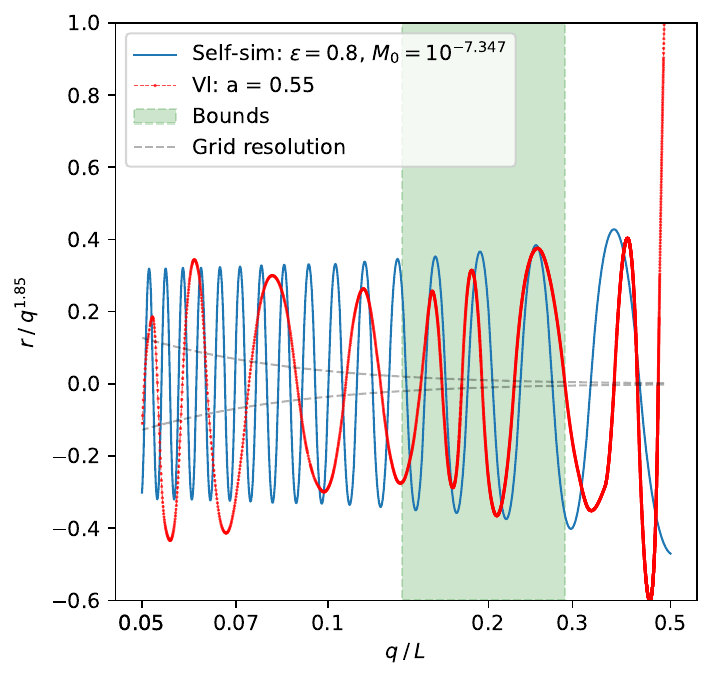}
    \includegraphics[width = 0.33\textwidth]{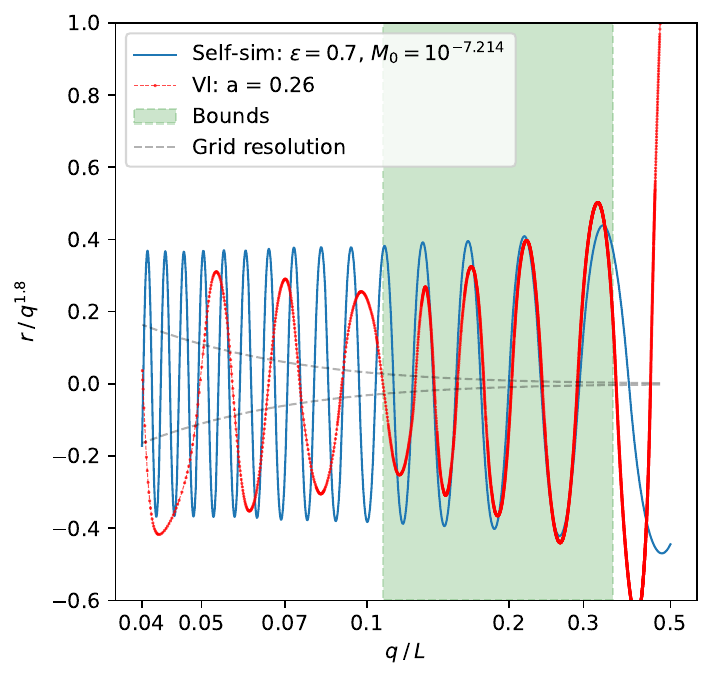}
    \includegraphics[width = 0.33\textwidth]{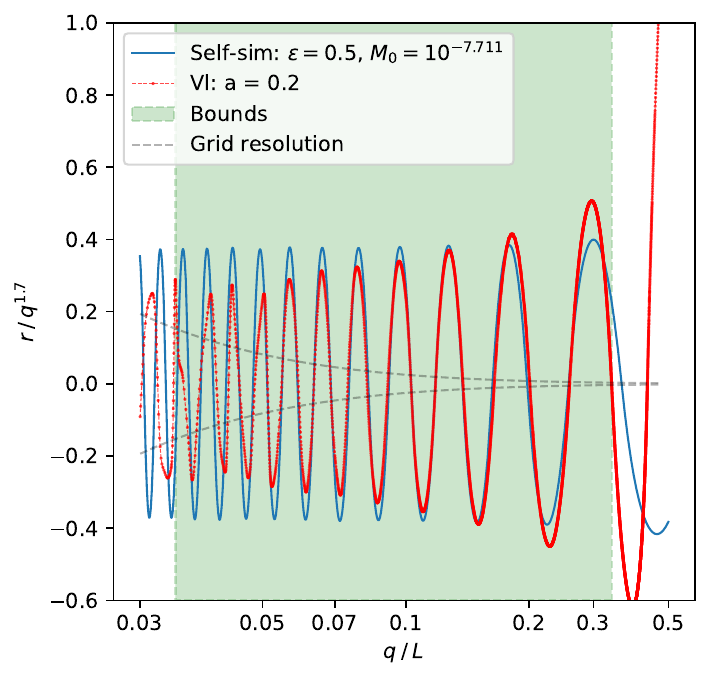}

    \includegraphics[width = 0.33\textwidth]{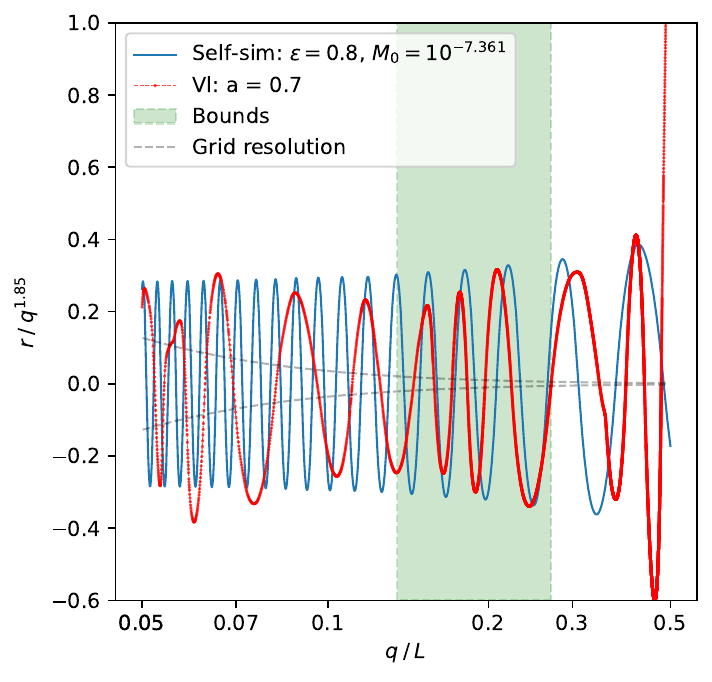}
    \includegraphics[width = 0.33\textwidth]{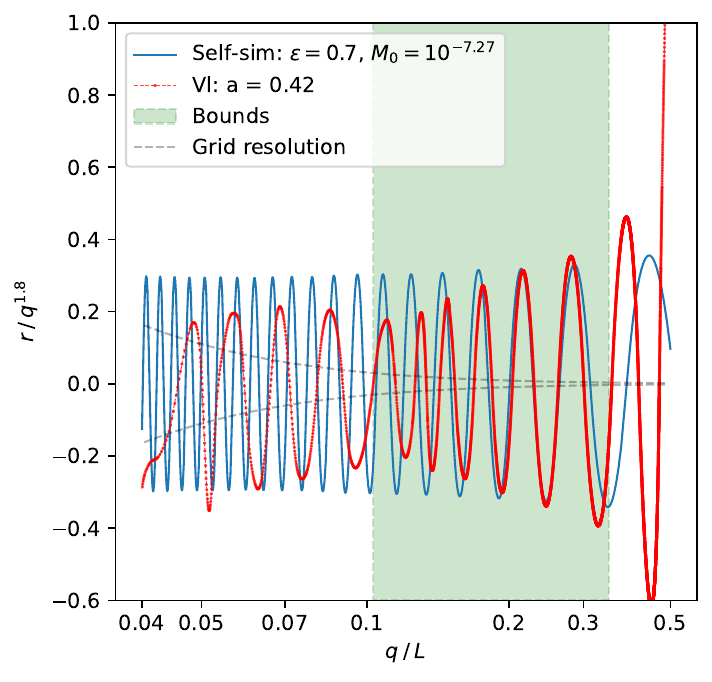}
    \includegraphics[width = 0.33\textwidth]{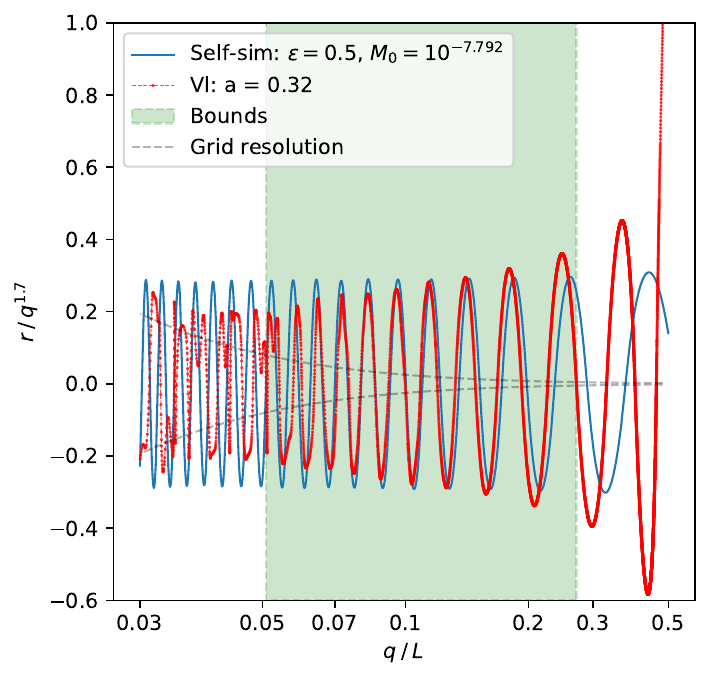}
    
    \centering
    \caption{$r-q$ curves of Lagrangian slices along $x$-axis for the three simulations - Q1D (left), ANI (middle), SYM (right) in increasing order of expansion factor from top to bottom. The data points (red, labelled 'Vl' as in Vlasov solver used for the simulations) have been cubically interpolated. The self-similar fits (blue) are made using the least-squares method with the parameter $\epsilon$ fixed at 0.8, 0.7 and 0.5 for Q1D, ANI and SYM simulations respectively. The subsets of particles (green) that follow the self-similar fits were computed assuming a threshold: $\Delta r/r_{\rm data} \le 10\%$. The axes have been scaled to make the features at lower values of $q$ more prominent. The spatial resolution, $\sim 0.0005$ comoving box units, is shown in black dashes.}
    \label{fig:vl_comp_r_q}
\end{figure*}

There are two ways to interpret the self-similar solution for $\lambda(\tau)$. Refer eq. \eqref{eq:tau}. If $t$ is varied while $M_{\rm i}$ is fixed, $\tau$ acts as time elapsed for a particle, with initial position $\mathbf{q}$ on a shell enclosing an initial mass $M_{\rm i}$, which oscillates about the center. In this case, the self-similar solution $\lambda-\tau$ can be scaled using eqs. \eqref{eq:tau} and \eqref{eq:lambda} to get the predicted $r-t$ trajectory of the corresponding Lagrangian point $\mathbf{q}$ on the phase-space sheet in our simulation. The initial mass profile $M_{\rm i}(q)$ used to map to the Lagrangian point $\mathbf{q}$ is measured from the Vlasov simulation.

Figure \ref{fig:vl_comp_r_t} shows self-similar fits (blue) to the position-time trajectories (red) of selected Lagrangian points along $x$ axis in SYM simulations. The fits are made using the standard least squares method. The data points are then cubically interpolated and compared with the fits to identify the points beyond which the relative differences between the data curves and the fits exceed 10\%. The time intervals within these bounding points are shown in green. Note that in each case, the trajectory follows the fit only after 1-2 oscillations after shell-crossing. A possible explanation could be that at $t_{\rm i}$, there is no collapsed material and the initial perturbation is not a power-law as assumed in the self-similar model (see eq. \eqref{eq:init_delta_ss}). It takes a brief period of relaxation for a power-law profile to build up before the trajectories approach a self-similar solution. %A point closer to the center relaxes and subsequently 'turns self-similar' earlier, the lower bound in $t$ is lower for lower values of $q$, refer fig. \ref{fig:q_bounds}.
The trajectories deviate again after tracing the fits for 3-4 oscillations, the reason for which could not be verified due to the sparsity of available snapshots at late times.

In the second interpretation, if $M_{\rm i}$ is varied keeping $t$ fixed, $\tau$ acts as a label running over particles in the snapshot $t$. The self-similar solution $\lambda-\tau$ can be scaled using eqs. \eqref{eq:tau} and \eqref{eq:lambda} to get the predicted locus of positions $r$ of Lagrangian points $\mathbf{q}$ in the snapshot $t$. The initial mass profile $M_{\rm i}(q)$ measured from the Vlasov simulation is used to map $\tau$ to the Lagrangian points $\mathbf{q}$. This interpretation has an advantage over the former - the Lagrangian points $\mathbf{q}$ can be sampled down to the grid resolution whereas the no. of snapshots we have are limited and sparse.

Figure \ref{fig:vl_comp_r_q} shows self-similar fits (blue) to the $r-q$ curves (red) of Lagrangian slices along $x$ axis for Q1D, ANI and SYM simulations in increasing order of expansion factor. The fits were made keeping $\epsilon$ fixed, allowing only $M_0$ to vary and then repeated for several values of $\epsilon$. For the purpose of presentation, $\epsilon = 0.8, 0.7$ and $0.5$ for Q1D, ANI and SYM respectively were chosen as representative values. Further details regarding the distribution of best-fit parameters are in section \ref{sec:param_dist}. The subsets of particles whose positions differ by $\le 10\%$ relative to the fits are shown in green. Similar to what we observed in the former approach, in each of the three simulations, the particles initially deviate for $\sim$ 1-2 oscillations after shell-crossing, then follow the fit for 3-4 oscillations in Q1D, 5-6 oscillations in ANI and 8-9 oscillations in SYM simulations and eventually, deviate again. Put differently, the subset of particles that follow the FG self-similar fit (green), increases in the order Q1D < ANI < SYM and it is clearly due to the lower bound decreasing in the same order. This suggests that the reason for the eventual deviation is correlated to the extent of non-radial dynamics arising from symmetry and not an artifact of numerical simulations. As expected, the FG solutions work best for SYM simulations where the trajectories are highly radial. The ANI and Q1D cases have increasingly elliptical collapses and transverse motion which lead to greater deviations. From the last available snapshots (bottom row), we may concur that the initial conditions do leave their imprint on the particle dynamics for 10-11 shell crossings at the very least.

Also, note the increase in frequency of oscillations from Q1D to SYM case. Consider the Lagrangian point $q\,/\,L = 0.1$ as an example. It has undergone 3 oscillations in Q1D, 5 oscillations in ANI and 6 oscillations in SYM case by $a = 0.32$. This is the reason why the best-fit self-similar solutions have correspondingly lower $\epsilon$ across the 3 simulations, even though we have the same mass in all of them. 

Each halo ought to be characterised by a single set of ($M_0, \epsilon$) in the FG model. What we find in our tests, is rather a distribution of best-fit parameters at different times and spatial regions of each halo, which we examine in section \ref{sec:param_dist}.
 
\subsection{Deviations}
\label{subsec:deviations}

\begin{figure}[]
    \centering
    \includegraphics[width = 0.45\textwidth]{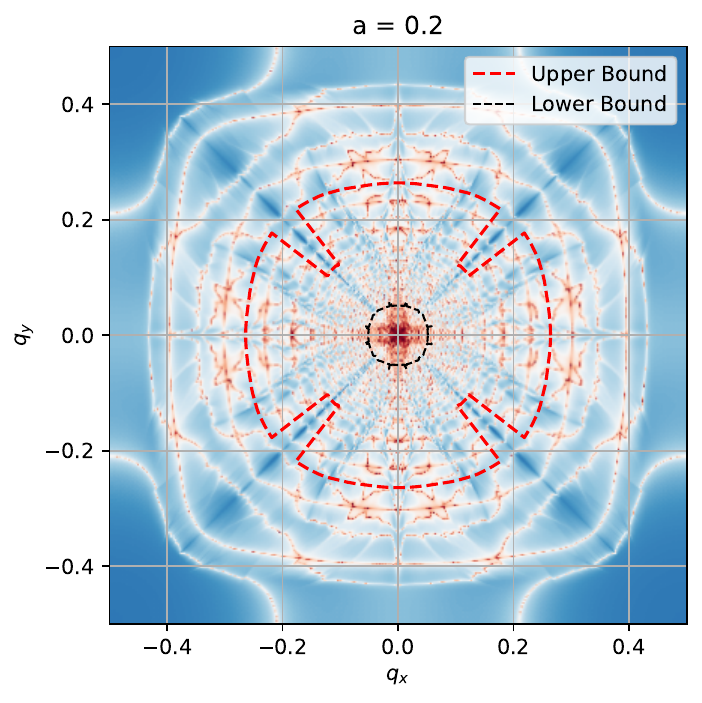}
    \caption{Bounds in Lagrangian space of the SYM simulation at the snapshot $a = 0.2$. The space has been color-mapped to corresponding Lagrangian density $\sim \left| \partial \mathbf{x} / \partial \mathbf{q} \right|^{-1}$. The contours of high density represent the caustics.}
    \label{fig:q_bounds_2D}
\end{figure}

\begin{figure*}
    \centering
    \includegraphics[width = 0.33\textwidth]{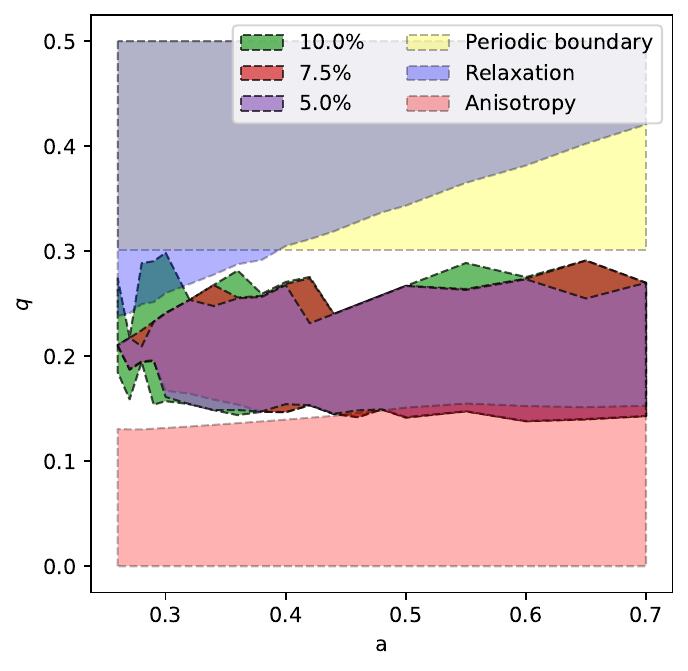}
    \includegraphics[width = 0.33\textwidth]{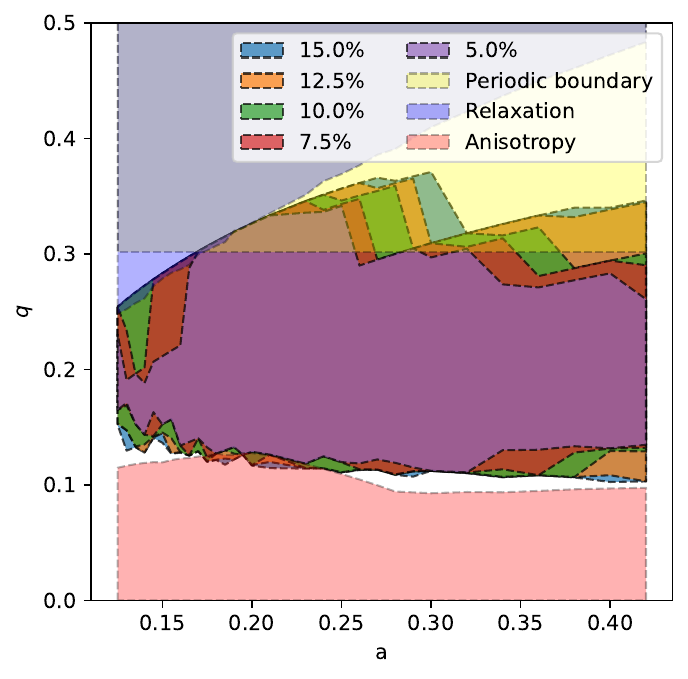}
    \includegraphics[width = 0.33\textwidth]{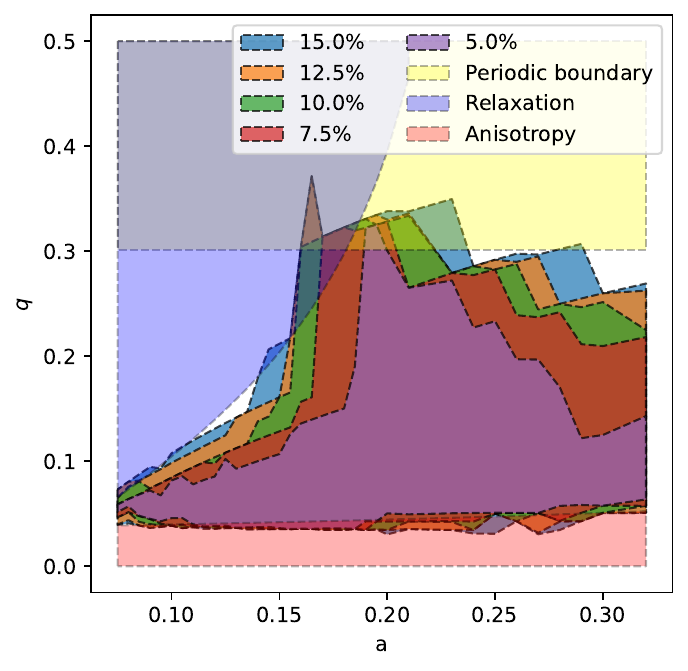}
    \caption{Evolution of bounds in Lagrangian space computed using $r -q$ curves of Lagrangian slices along $x$ axis for three cases: Q1D (left), ANI (middle), SYM (right). The parameter $\epsilon$ was fixed at 0.8, 0.7 and  0.5 respectively. The multicolored patches labelled with percentages depict the subset of particles in lagrangian space for which the relative difference in their positions between the data curves and the self-similar fits are within thresholds of 5\%, 7.5\%, 10\%, 12.5\% and 15\%. The spaces excluded by bounds arising from relaxation, periodic boundaries and anisotropy have been shown in blue, yellow and red patches respectively.}
    \label{fig:q_bounds}
\end{figure*}

In this subsection, we study the deviations of particle trajectories from the FG self-similar fits in greater detail. In particular, we look at where the deviations occur, and how they change with time as well as across the three simulations to try and identify the causes behind them.

The subset of particles bounded in green denotes the portion of the phase-space sheet whose eulerian position in the simulation is matched by the self-similar fits within a 10\% error. It implies that an entire halo does not 'turn self-similar' as a whole. By repeating this fitting exercise for Lagrangian slices along different directions $\theta$ for a given snapshot of a simulation, we find the subset of particles to be confined between two concentric rings in Lagrangian space $q_x - q_y$ whose radii are equal to the computed bounds. Figure \ref{fig:q_bounds_2D} depicts the bounds in Lagrangian space of the SYM simulation at one of the snapshots as an example.

Fig. \ref{fig:q_bounds} shows the evolution of the bounds computed for Lagrangian slices along the $x$ axis in each of the three simulations. Note that the first snapshot wherein a self-similar fit could converge is always after shell-crossing along $y$-axis and at increasingly late times $a_r = 0.075, 0.125, 0.26$ going from SYM to ANI to Q1D simulations (refer Table \ref{table:vl_sim_params}). As discussed earlier, a power-law profile needs to build up before the trajectories can approach a self-similar solution. The lower amplitude of initial displacement $\epsilon_y$ delays the shell-crossing along $y$ axis and as a consequence, relaxation and profile build-up are also delayed. 

Looking at the subset of particles bound in green, the upper bound grows initially (clearly observed in ANI and SYM cases), but then stagnates, even reducing afterward. To track the initial growth, we hypothesize that after shell-crossing, the particles take 1-2 oscillations to relax to the self-similar fits. Relaxation \citep{Lynden_bell_1967} refers to the process of phase-space mixing following which particle motion can be described using acceleration computed from a smooth potential field. The subset of such particles grows as more particles starting farther away from the center (at higher $q$) gradually relax and turn self-similar. This argument is tricky for the Q1D and ANI cases, as they have different shell-crossing times along the $x, y$ axes, whereas the FG model just has one owing to circular symmetry. In the simulations, we observed that the particles begin to relax to self-similar fits only after the last shell-crossing has occurred (along the $y$-axis in our cases). Since we cannot track the two shell-crossings separately using the FG model, we simply use the shell-crossing from theory (which might not correspond to the actual shell-crossing along $y$-axis) to build our argument. Referring to the top left panel of fig. \ref{fig:self_sim_sol}, 1-2 oscillations after shell-crossing roughly corresponds to $\tau = 7-20$ for $\epsilon = 0.8$, $\tau = 6-18$ for $\epsilon = 0.7$, $\tau = 5-13$ for $\epsilon = 0.5$ in Q1D, ANI and SYM cases respectively. Using eq. \eqref{eq:tau}, the $M_i$'s corresponding to these values of $\tau$ were computed, which were then mapped to the bounding $q$'s in Lagrangian space for each snapshot. The subsets of particles that lie above these bounds are shown in blue and they seem to coincide well, at least for the SYM case, with the particles that initially deviate.

As all particles would eventually relax, the upper bound should have kept increasing until $q\,/\,L = 0.5$, however, it stagnates for Q1D and ANI cases and decreases for the SYM case after it reaches $q \sim 0.3$. Taking a look at the $r-q$ curves in figure \ref{fig:vl_comp_r_q} at the last available snapshots (bottom row), we notice that the particles with $q/L$ roughly $\ge 0.3$ have not completed as many oscillations as they ought to if they had traced the self-similar fits. This is clearly due to a difference in the force between FG theory which assumes a single isolated halo and our simulations which have periodic boundaries. As a hand-waving argument, consider the motion of a particle with the same initial position $q$ in the two setups, one with a single isolated halo at the center and the other having an additional halo at $x\,/\,L = 1$. Initially, their relative difference in positions is negligible but grows gradually. Thus, the initial deviation from self-similarity is dominantly due to the time taken for relaxation, after which the deviation caused by periodic boundaries takes over once the relative difference in positions crosses the \% threshold we set to compute the bounds in $q$. The relative difference between the force fields in the two setups scales as $x/(L-x)$ ; $x/L \in [0.0, 0.5]$, which implies that if the initial position $q$ was closer to the boundary, the particle would suffer from a greater erroneous force arising from the halo image. Hence, it would accumulate error in its trajectory faster and deviate from self-similar fit earlier. This explains why the upper bound stagnates and moves to lower $q$ later in time, most prominently observed for the SYM case. The subsets of particles that experience a relative error in force $\ge 40\%$ are shown in yellow in fig. \ref{fig:q_bounds}.

The lower bound remains roughly constant over time in each simulation, but its magnitude decreases going from Q1D to ANI to SYM. Since the upper bound stagnates at $q/L \sim 0.3$ for each simulation, it implies that the subset of particles in agreement with FG solutions increases in the same order. We hypothesize this to be correlated to the extent of transverse motion across the simulations. Figure \ref{fig:trans_motion} shows that the degree of transverse motion is indeed the highest for particles in Q1D simulation. Since FG solutions assume fully radial orbits, the fits for $r-q$ curves of Lagrangian slices along non-axial directions barely converge in Q1D and ANI simulations. In figure \ref{fig:beta_param}, the transition of the anisotropy parameter $\beta$ from $1$ to $0$ as we move closer to the halo center roughly demarcates the halo into two regions with different dynamics - the outer region, dominated by radial infall and the inner region, where the velocity distribution approaches isotropy after violent relaxation due to transverse motions. For Lagrangian slices along axial directions, we could expect to see deviations from FG solutions once the amplitude of oscillations falls within the radius of the inner region $r_{\rm trans}$ where the transverse motion is no longer negligible. From fig. \ref{fig:vl_comp_r_q}, we note that the asymptotic amplitude of oscillations can be parameterised by a power law: $r/aL = A(a) q^{\alpha}$, where $\alpha = 1.85, 1.8, 1.7$ for Q1D, ANI and SYM cases respectively and $A(a) \sim 0.3 - 0.5$. The transition radius $r_{\rm trans}(a)$ was computed assuming $\beta({r_{\rm trans}}) = 0.5$ using sigmoid fits to the anisotropy profile at each snapshot. The subset of particles whose amplitudes of oscillations are less than $r_{\rm trans}$ at a given snapshot i.e. $A(a) q^{\alpha} \le r_{\rm trans}(a)$ are shown in red in fig. \ref{fig:q_bounds}.

Summing up the deviations seen in particle trajectories:
\begin{itemize}
    \item Particles typically take 1-2 oscillations after the last shell-crossing (along $y$-axis) to relax to self-similar motion, resulting in the initial upper bounds. Therefore, this deviation is due to physical reasons.

    \item On the other hand, the upper bounds in the later stages are due to the periodic boundary condition that exerts artificial forces on the particles. The difference between FG trajectories and simulated trajectories of particles gradually increases, faster for the particles which start closer to the boundaries. This deviation is an artifact of our simulations.

    \item Finally, the lower bound is associated with the fact that we are comparing FG self-similar fits assuming fully radial motion to simulated trajectories in halos that have significant transverse motion in their central regions. Once the amplitude of oscillations of the particles decreases down to the radius inside which transverse motion turns significant, they start to deviate. It is to be expected that the lower bound decreases across the simulations in increasing order of circular symmetry i.e. the SYM simulations are the most consistent with the FG model of self-similarity. It would therefore be interesting to consider more general models of self-similarity which take transverse motion as well as elliptical collapse into account.
\end{itemize}

\subsection{Mass and density profiles}
\label{subsec:mass_rho_profiles}

\begin{figure*}
    \includegraphics[width = 0.33\textwidth]{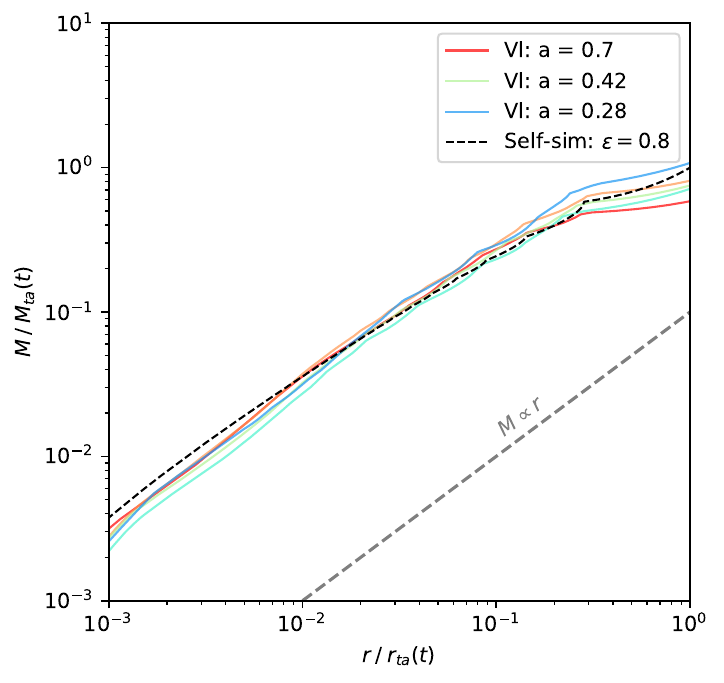}
    \includegraphics[width = 0.33\textwidth]{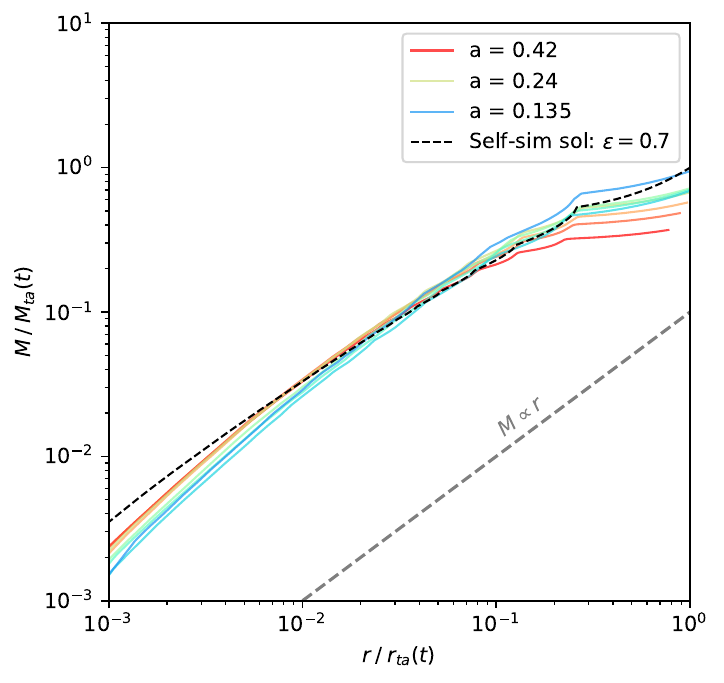}
    \includegraphics[width = 0.33\textwidth]{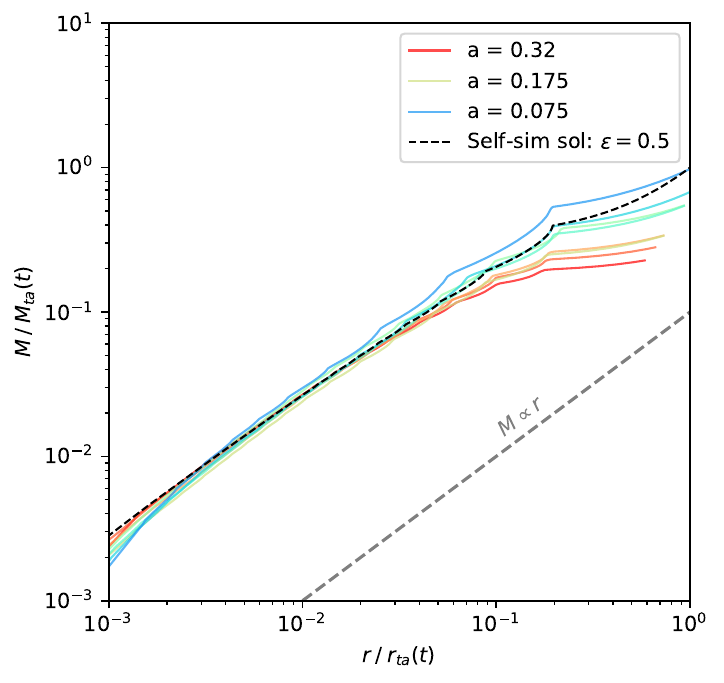}

    \includegraphics[width = 0.33\textwidth]{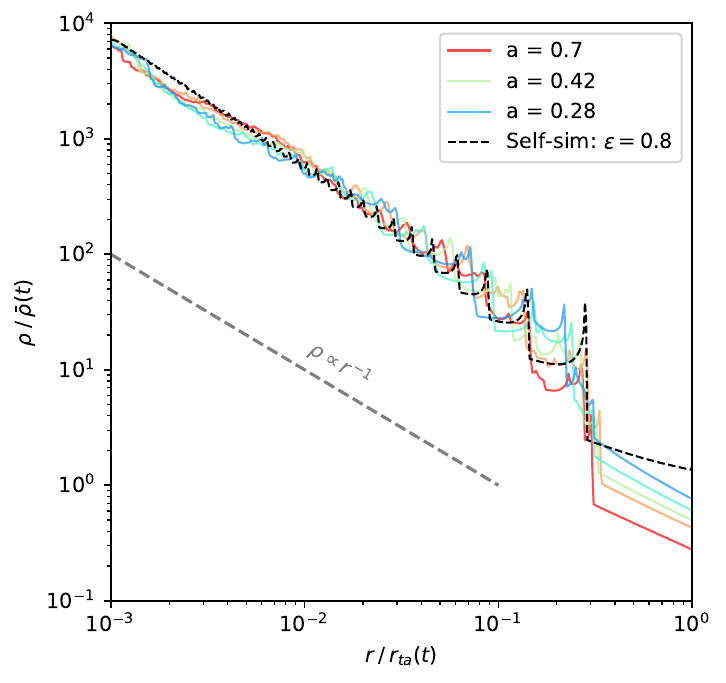}
    \includegraphics[width = 0.33\textwidth]{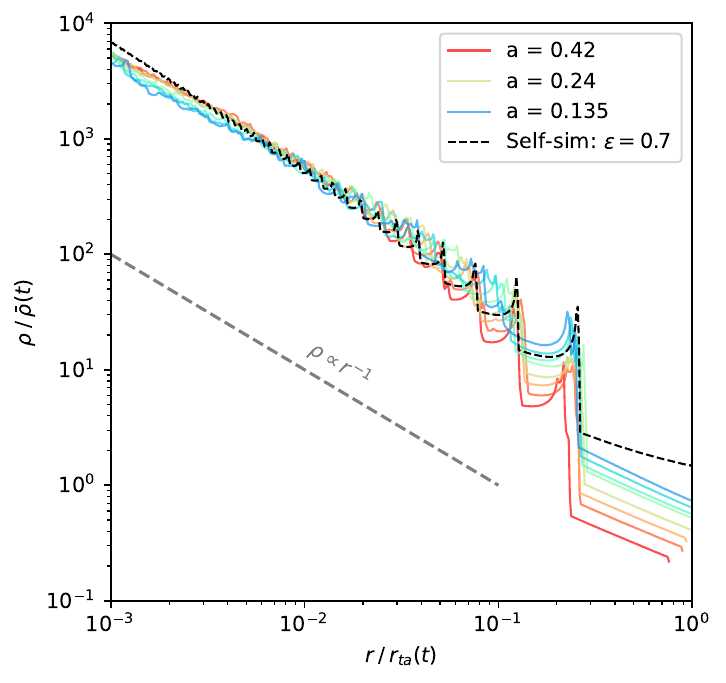}
    \includegraphics[width = 0.33\textwidth]{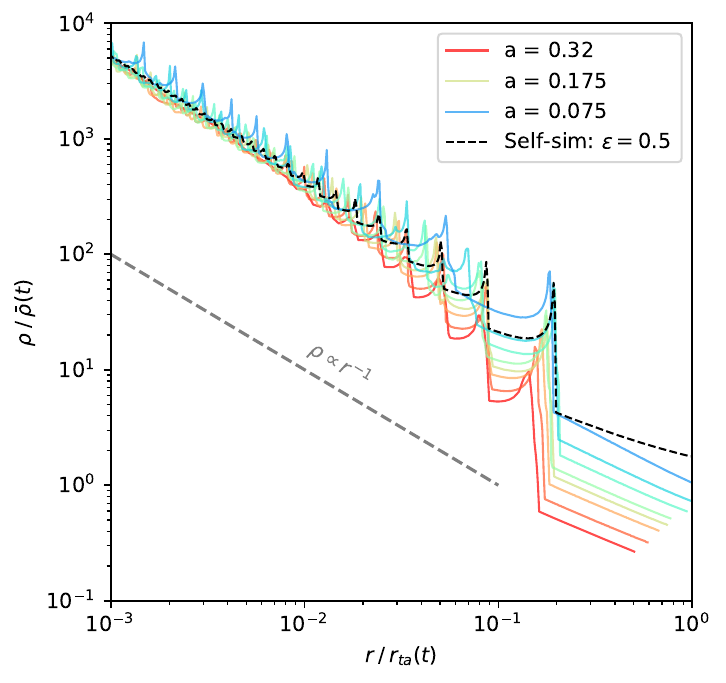}
    \centering
    \caption{ Mass (top) and density (bottom) profiles from the three simulations: Q1D (left), ANI (middle) and SYM (right) normalised w.r.t. to the mass $M_{\rm ta}$ and radius $r_{\rm ta}$ of the turnaround region using the best-fit $M_0(\epsilon)$ at different snapshots, color coded from blue to red in increasing order of expansion factor. The black dashed curve denotes the self-similar predictions for the mass and density profiles corresponding to $\epsilon = 0.8, 0.7, 0.5$ for Q1D, ANI, SYM respectively.}
    \label{fig:vl_comp_mass_density_profile}
\end{figure*}

After analysing the motion of particles in our simulations, we turn to mass and density profiles. The self-similar prediction for mass profile inside turnaround region $\mathcal{M}(r/r_{\rm ta})$ is obtained by iteratively solving eqs. \eqref{eq:main_eq_lambda} and \eqref{eq:main_eq_mass_simp}, from which we can derive the density profile using eq. \eqref{eq:density_profile}. These are, again, characterised by $M_0$ and $\epsilon$. From the simulations, we measure the cumulative mass and circularly averaged density in radial bins. Using least squares, we obtain the best-fit $M_0$ for each snapshot, keeping $\epsilon$ fixed at 0.8, 0.7 and  0.5 for Q1D, ANI and SYM respectively. Figure \ref{fig:vl_comp_mass_density_profile} shows the self-similar solution overplotted with the mass and density profiles from the three simulations normalised w.r.t. the mass $M_{\rm ta}$ and radius $r_{\rm ta}$ of the turnaround region using the best-fit $M_0(\epsilon)$ at multiple snapshots, color coded from blue to red in increasing order of expansion factor. Note that the first snapshot for which self-similar fit could converge to the mass (or density) profile is at increasingly late times $a_M = 0.055, 0.1, 0.2$ going from SYM to Q1D, while satisfying $a_{{\rm SC},y} \le a_M \le a_r$ in each case (refer table \ref{table:vl_sim_params}). Non-convergence of self-similar fit means that a power-law profile has not been fully built up yet. This verifies that lower $\epsilon_y$ delays the shell-crossing along $y$, which delays relaxation and profile build-up, which in turn delays the approach of trajectories to self-similar solutions.

The magnitude and slopes of the mass and density profiles are in good agreement. The number and location of caustics (spikes in mass and density profiles) differ initially but eventually conform to the self-similar fits. This can be understood from the $r-q$ curves of Lagrangian slices in fig. \ref{fig:vl_comp_r_q}. Caustics correspond to the extrema of oscillations. As the self-similar fits to the number and amplitude of oscillations improve with time, so does the number and location of caustics.  At late times, however, we note a dip in the slope of mass and density profiles at outer region of the halos. This stems from the fact that the FG model assumes indefinite infall of mass onto the halo whereas in our simulations, once the mass $M_{\rm ta}$ (or radius $r_{\rm ta}$) of turnaround region grows beyond the mass (or box size) of our simulation, there is a deficit of infalling matter. The periodic boundaries holding back the particles closer to the boundaries also aggravate this issue.

The time-averaged slopes of the mass profiles in the range $0.001 \le r/r_{\rm ta} \le 0.01$ for Q1D, ANI and SYM cases are 1.05, 1.10 and 1.00 respectively. These are in good agreement with an asymptotic slope of 1.0 predicted in FG for cylindrical symmetry. Since all the simulations produce nearly the same slope despite differing in their values of best-fit $\epsilon$, it verifies that the slope is indeed independent of $\epsilon$ (in turn, the halo mass) for the 2D case.

\subsection{Scale-free space}
\label{subsec:scale_free_space}

\begin{figure*}[]
    \centering
    \includegraphics[width = 0.33\textwidth]{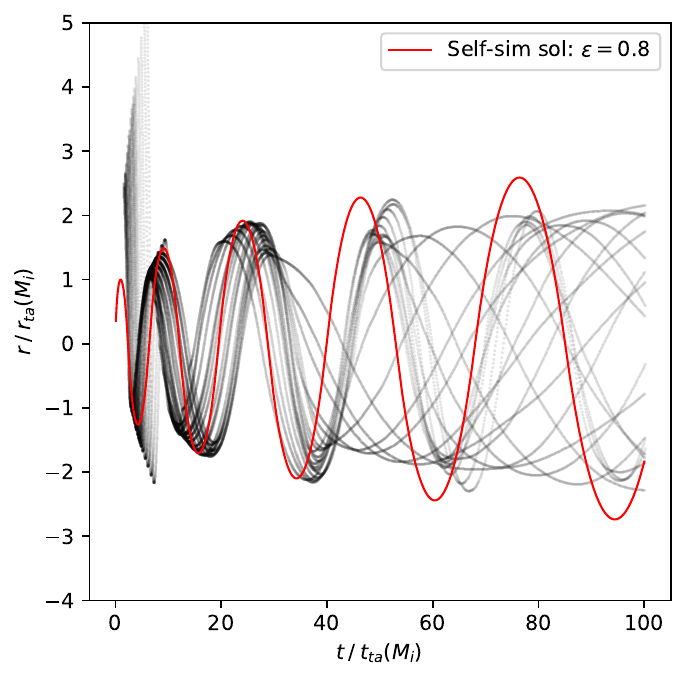}
    \includegraphics[width = 0.33\textwidth]{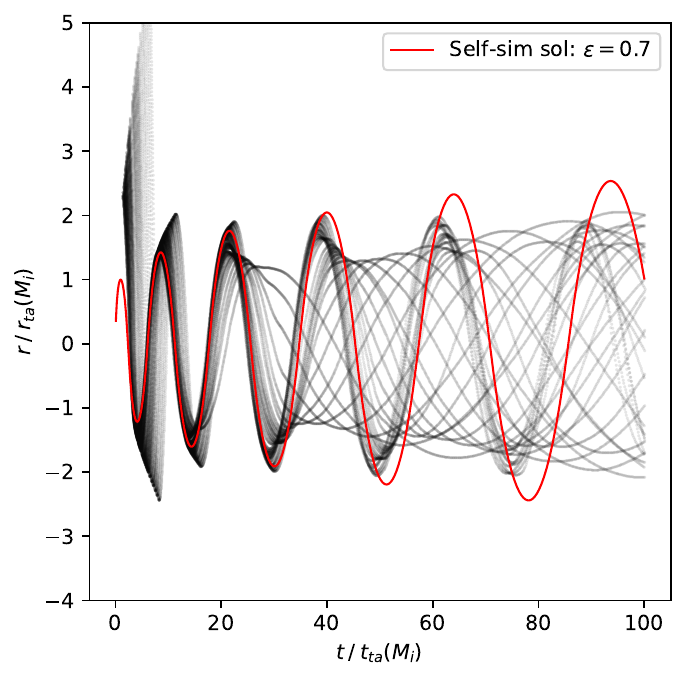}
    \includegraphics[width = 0.33\textwidth]{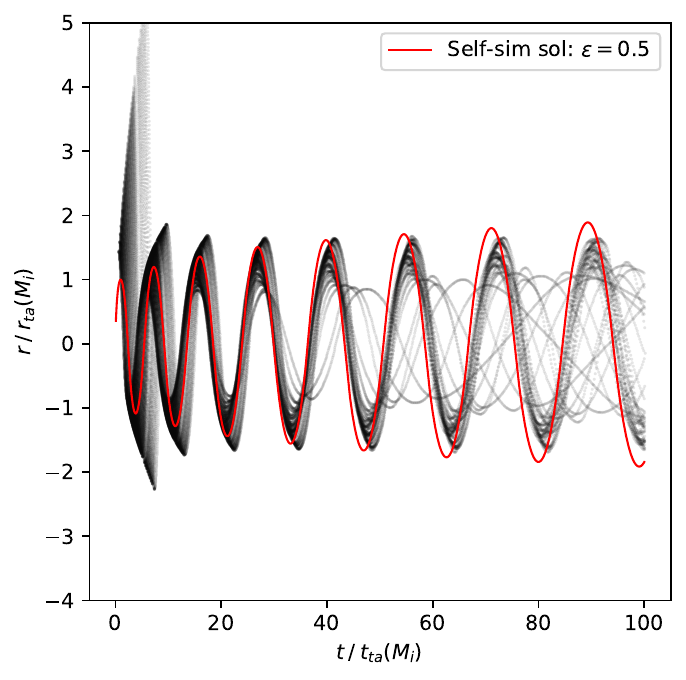}

    \includegraphics[width = 0.33\textwidth]{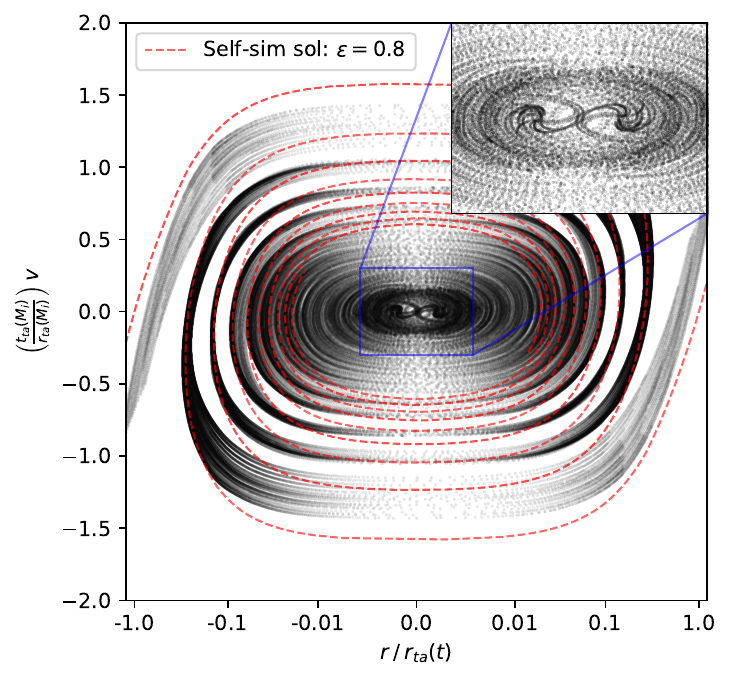}
    \includegraphics[width = 0.33\textwidth]{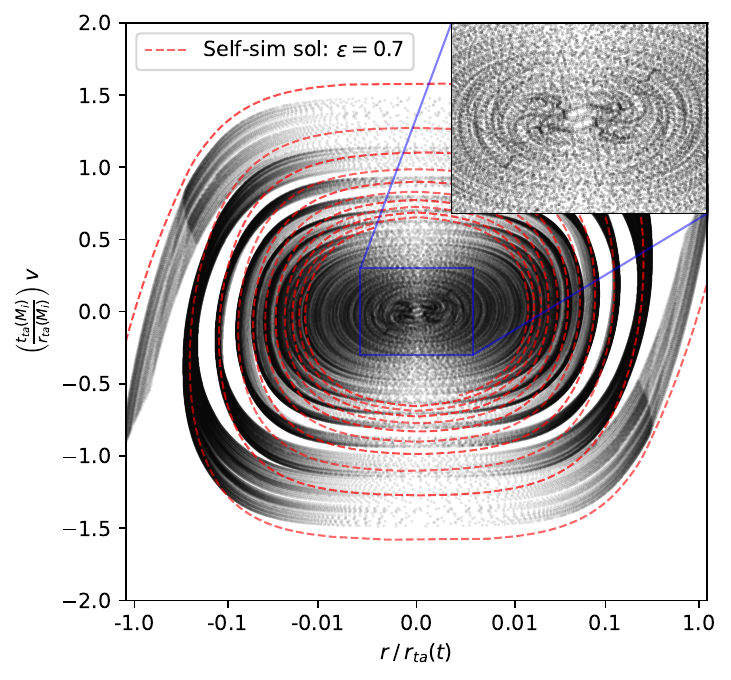}
    \includegraphics[width = 0.33\textwidth]{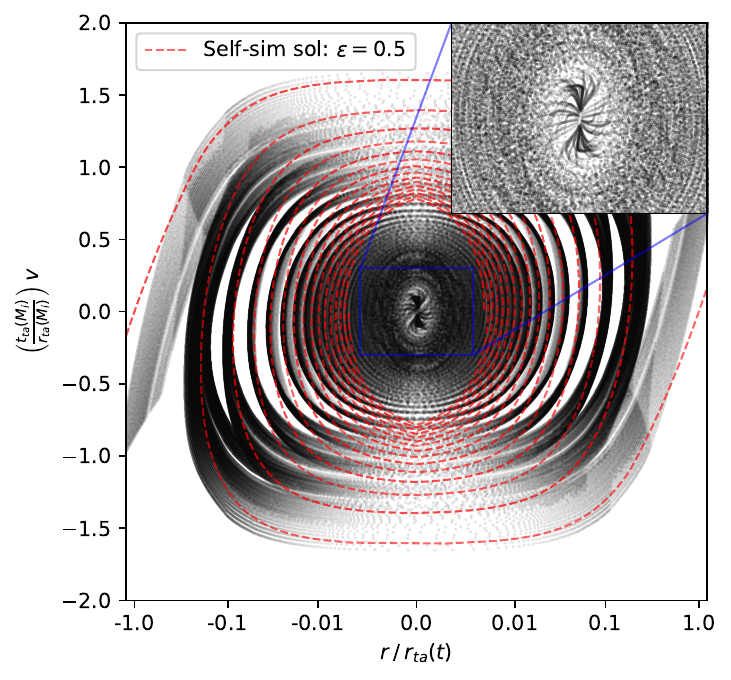}
    
    \caption{Superposition of position-time (top) and phase-space (bottom) trajectories of particles of Lagrangian slices along $x$ axis from all the snapshots (where a fit could converge) after being scaled w.r.t turnaround computed using equations \eqref{eq:tau} and \eqref{eq:lambda} using the best fit $M_0(\epsilon)$ obtained from fitting $r-q$ curves (refer fig. \ref{fig:vl_comp_r_q}) in sec. \ref{subsec:part_traj}. Zoomed panels of the central regions in phase space have also been added. Self-similar solutions for $\epsilon = 0.8, 0.7, 0.5$ corresponding to Q1D (left), ANI (middle) and SYM (right) are shown in red.}
    \label{fig:vl_comp_self_sim_space}
\end{figure*}

Self-similarity implies that trajectories in position and phase space basically have the same shape. Upon scaling w.r.t. the characteristic length and time - the turnaround radius and time in the FG model, the trajectories should overlap. As a consistency check, after obtaining the best-fit parameter $M_0(\epsilon)$ using fits to particle trajectories in sec. \ref{subsec:part_traj}, we scale the trajectories w.r.t to turnaround using eqs. \eqref{eq:tau} and \eqref{eq:lambda}, and overplot all the snapshots together with the self-similar solutions for $\epsilon = 0.8, 0.7, 0.5$ for Q1D, ANI, SYM cases respectively in figure \ref{fig:vl_comp_self_sim_space}.

The superposition of scale-free position-time trajectories from simulations, shown in the top row of panels, is strikingly good. The curves corresponding to later snapshots follow the self-similar curves longer in each simulation. As expected, the SYM simulations are the most consistent, with most of the curves being in agreement for roughly 7-8 oscillations. We also note that the oscillation frequency in each curve decreases with $\tau$ i.e. more the no. of oscillations a particle completes, less is its oscillation frequency. The $r-q$ curves for snapshots at later times would thus be better fit by self-similar solutions corresponding to higher values of $\epsilon$ (refer to the top-left panel of fig. \ref{fig:self_sim_sol}). This is indeed what we found upon redoing the fits for different values of $\epsilon$, discussed in detail in section \ref{sec:param_dist}. The parameter $\epsilon$ is inversely related to the mass accretion rate, which means that the mass accretion rate in our simulations is less compared to the expectation from the FG model at later snapshots. This is clearly due to the deficit of infalling matter in our simulations, which also led to the dips in mass and density profiles at larger radii at later snapshots. 

In the phase space curves shown in the bottom row of panels, the self-similar spiral pattern is remarkably consistent across the snapshots for each simulation. It is crucial to note that even though the phase-space spirals in Q1D and ANI show deviations from FG self-similar solutions, the simulated curves superpose quite well within themselves. This insinuates the existence of a homothetic transform (or self-similar solution) more general than the FG solution alone, one that incorporates transverse motion. In the zoomed panels of the central regions, we can clearly see the resonant modes compromising the mean-field limit, especially in Q1D. This is a numerical defect. The radius below which the motion of particles is contaminated by these resonant modes is approximately $r/aL \sim 1-2 \, \times \, 10^{-3}$, which is greater than the grid resolution $\sim 5 \times 10^{-4}$. However, this numerical bound is still less than the physical lower bound arising from transverse motion below the transition radius, the least of which is for the SYM case: $r_{\rm trans}/aL \ge 5 \times 10^{-3}$ (refer fig. \ref{fig:beta_param} and subsection \ref{subsec:deviations}).

\subsection{Transverse motion and anisotropy parameter}
\label{subsec:trans_beta}

\begin{figure*}[]
    \centering
    \includegraphics[width = 0.33\textwidth]{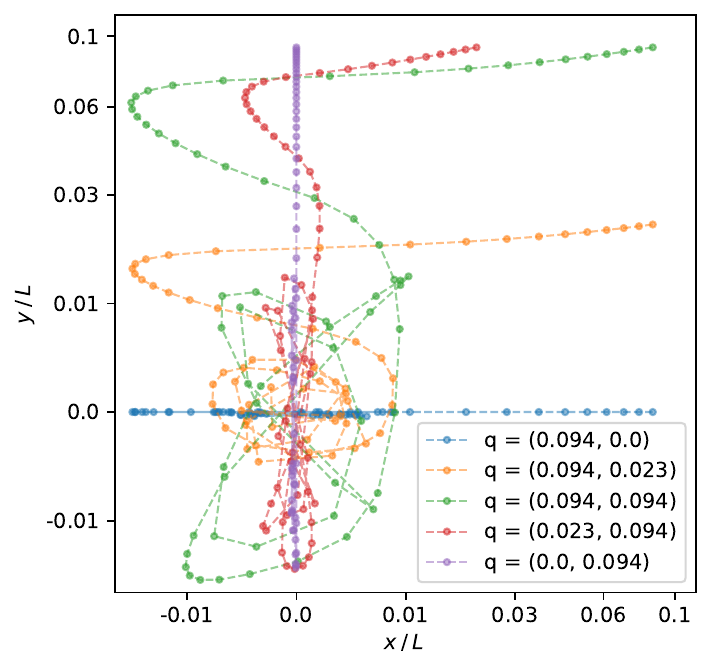}
    \includegraphics[width = 0.33\textwidth]{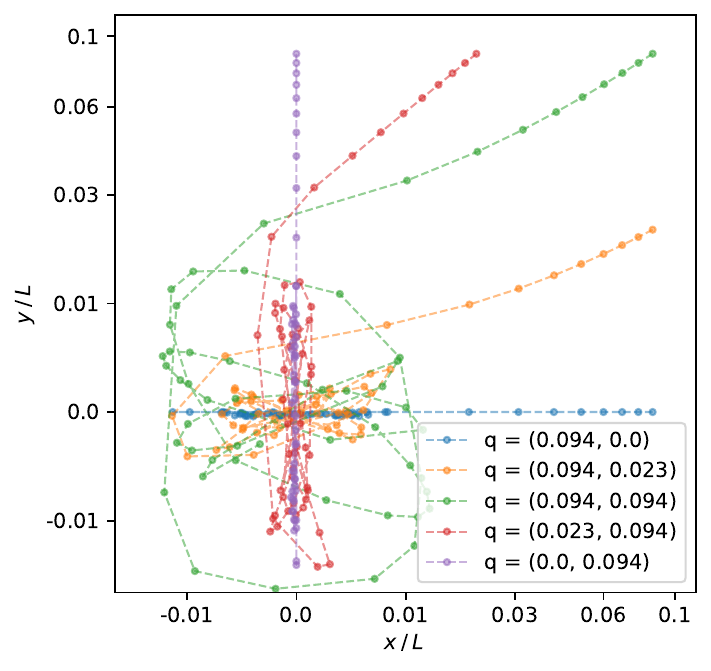}
    \includegraphics[width = 0.33\textwidth]{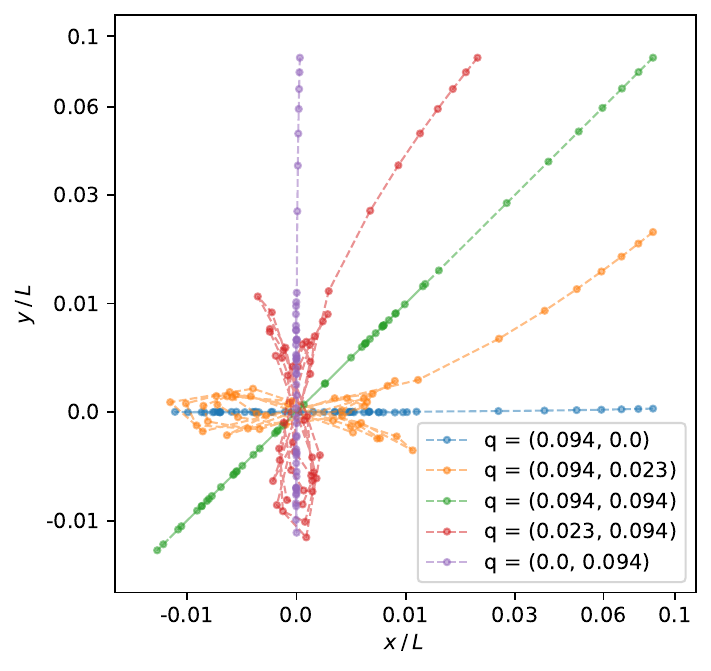}
    \caption{$x$-$y$ trajectories of five Lagrangian points $(q_x/L, q_y/L) = (0.094, 0.0), (0.094, 0.008), (0.094, 0.094), (0.023, 0.094), (0.0, 0.094)$ for the 3 cases - Q1D (left), ANI (middle), SYM (right) . The axes have been logarithmically scaled to feature the motion close to the center prominently. }
    \label{fig:trans_motion}
\end{figure*}

\begin{figure*}[]
    \centering
    \includegraphics[width = 0.33\textwidth]{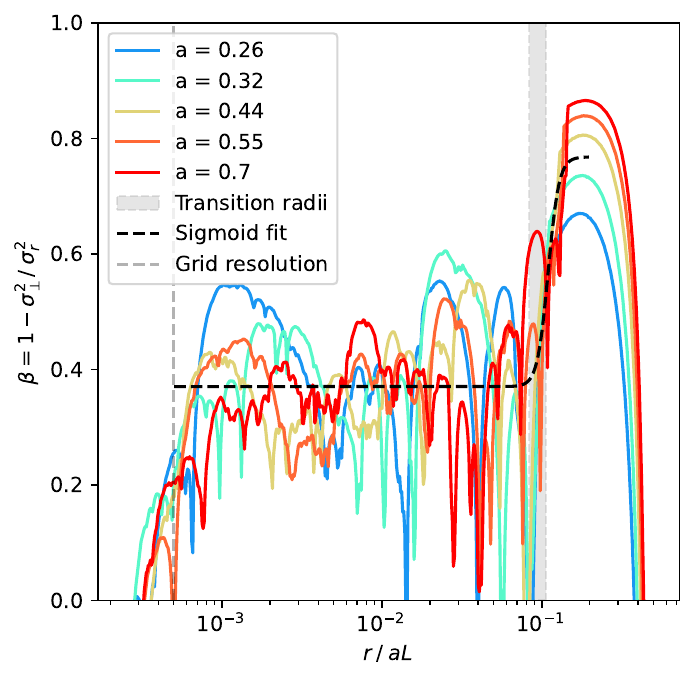}
    \includegraphics[width = 0.33\textwidth]{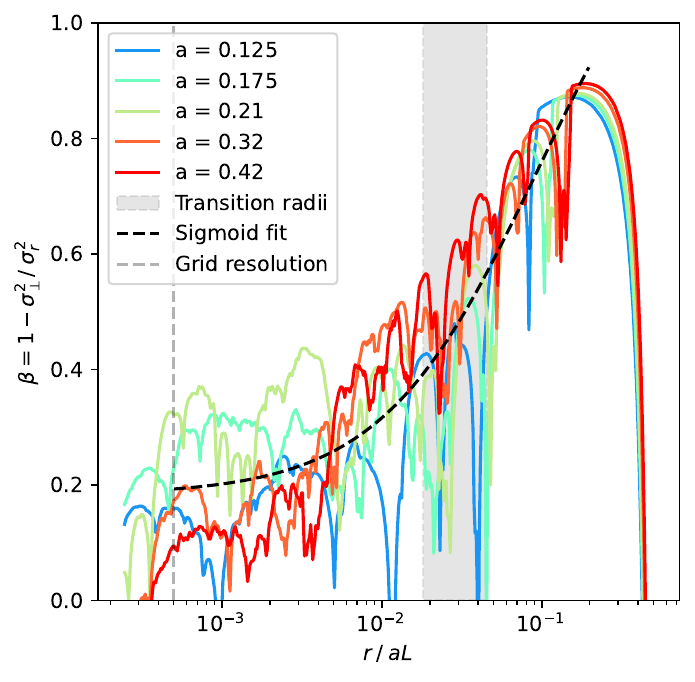}
    \includegraphics[width = 0.33\textwidth]{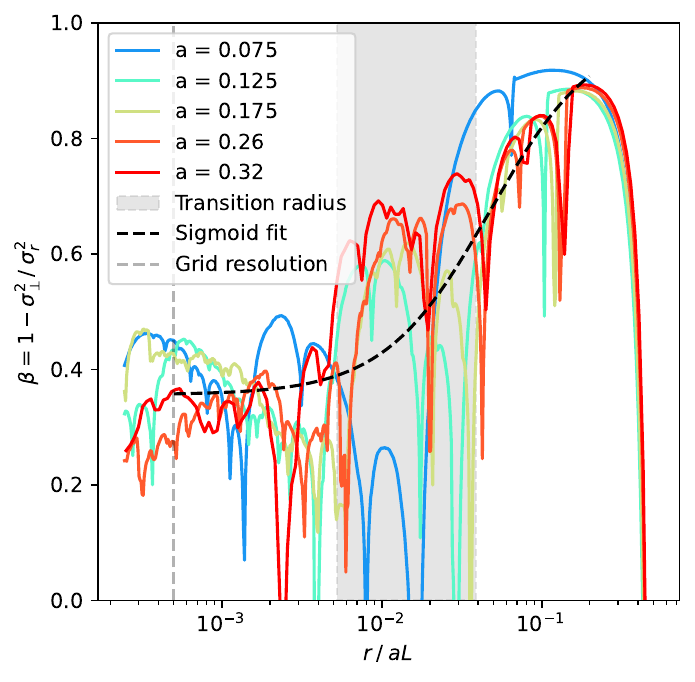}
    
    \caption{Radial profiles of anisotropy parameter $\beta = 1 - \sigma_{\perp}^2/\sigma_r^2$ at several snapshots for the 3 cases - Q1D (left), ANI (middle), SYM (right). The grid resolution $= 0.0005$ comoving box units is marked in grey dashed lines. The values of $r_{\rm trans}$ corresponding to $\beta = 0.5$ over different snapshots are marked as the grey patch. Illustrative sigmoid fits to the profiles are shown in black dashed lines.}
    \label{fig:beta_param}
\end{figure*}

To corroborate our hypothesis that non-radial dynamics cause particle trajectories in the interior of halos to deviate from FG solutions, we probe the extent of transverse motion and anisotropy in our simulations. 

Fig. \ref{fig:trans_motion} shows the $x-y$ trajectories of particles with initial positions along different directions in the three simulations. For the SYM case, only the particles starting along $x = 0, x = y, y = 0$ directions have completely radial trajectories. The other trajectories, while being mostly radial, do show deviations. For ANI and Q1D cases, only the particles starting along axial directions are radial. Q1D being the most asymmetrical has non-axial trajectories exhibiting the greatest extent of transverse motion.

To determine the region of the halos where transverse motion starts to be significant, we look into the anisotropy parameter which is defined as $\beta(r) = 1 - \sigma_{\perp}^2(r) / \sigma_r^2(r)$, where $\sigma_{\perp} , \sigma_r$ are the transverse and radial velocity dispersions respectively. For radial orbits, $\sigma_{\perp}^2 = 0 \implies \beta = 1$. For virialised orbits, $\sigma_{\perp}^2 = \sigma_r^2 \implies \beta = 0$. The higher the value of $\beta$, the more radial the particle trajectories and hence, we would expect better fits to FG solutions. Fig. ~\ref{fig:beta_param} shows the radial profile of the anisotropy parameter for the three simulations at several snapshots. The decrease of $\beta$ from 1 to 0 as we move inwards suggests that in the outer region, the dynamics is dominated by radially infalling matter, whereas in the inner region, due to isotropisation in velocity space, the particle trajectories have a non-negligible transverse component. To estimate the radius $r_{\rm trans}$ inside which transverse motion starts to be significant, sigmoid fits to the radial profile of anisotropy parameter were made for each snapshot and $r_{\rm trans}$ was determined using $\beta(r_{\rm trans}) = 0.5$. The fits were made in the range $0.0005 \ge r \, / \, aL \ge 0.2$. The grey regions denote the ranges of values of $r_{\rm trans}$ over all the snapshots in each simulation. The key observation is that the transition from $\beta = 1$ to $\beta = 0$ happens at correspondingly smaller radii the closer the simulation is to circular symmetry i.e. $r_{\rm trans}/aL \in [0.08, 0.1], [0.01, 0.04], [0.005, 0.02]$ for Q1D, ANI and SYM respectively. Also, note the unusually high value of $\beta$ at $r < r_{\rm trans}$ for Q1D. Since Q1D deviates the most from circular symmetry, one would expect greater isotropisation in velocity space close to the center and hence, lower $\beta$, which is not what we observe. One hypothesis could be that the Q1D particles indeed show a greater extent of transverse motion till they undergo the first crossing along the x-axis, after which, they move more or less along the y-axis and their collapse into the center of the halo (shell-crossing along the y-axis) is more radial than that of the ANI particles, leading to a higher value of $\beta$ in the central region. It does not seem to be a numerical artifact like the resonant modes since its extent $r_{\rm trans} / aL \sim 10^{-1}$ far exceeds the extent of the resonant modes $r / aL \sim 1-2 \, \times \, 10^{-3}$

In actual CDM cosmologies seeded by Gaussian random fields, the collapse of initially overdense perturbations leading to the formation of halos is better modelled by ellipsoidal instead of spherical collapse and the particles exhibit significant non-radial motion as well. Such halos would be closer in resemblance to the Q1D than the SYM case. Therefore, a self-similar analysis of 3D CDM halos would entail the inclusion of ellipsoidal collapse and transverse motion to build a more generic model of self-similarity.

%--------------------------------------------------------------------
\section{Parameter distribution}
\label{sec:param_dist}

\begin{figure*}
    \includegraphics[width = 0.33\textwidth]{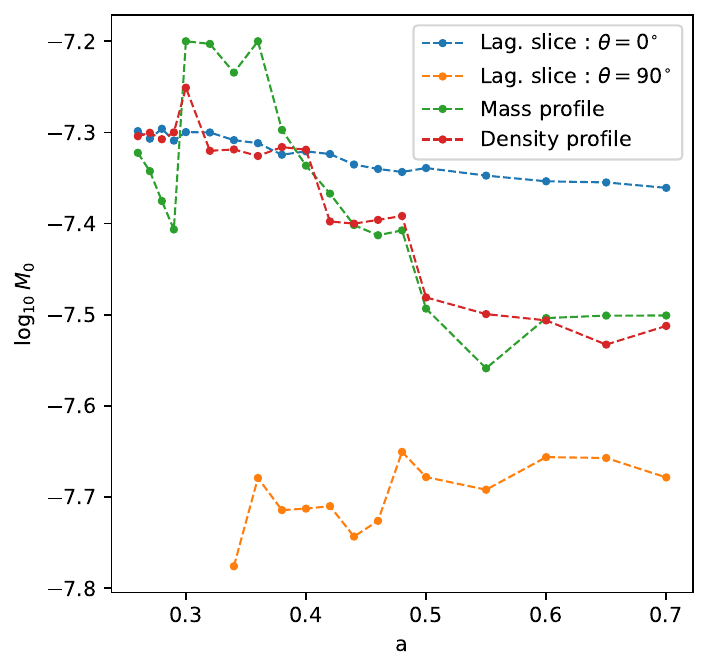}
    \includegraphics[width = 0.33\textwidth]{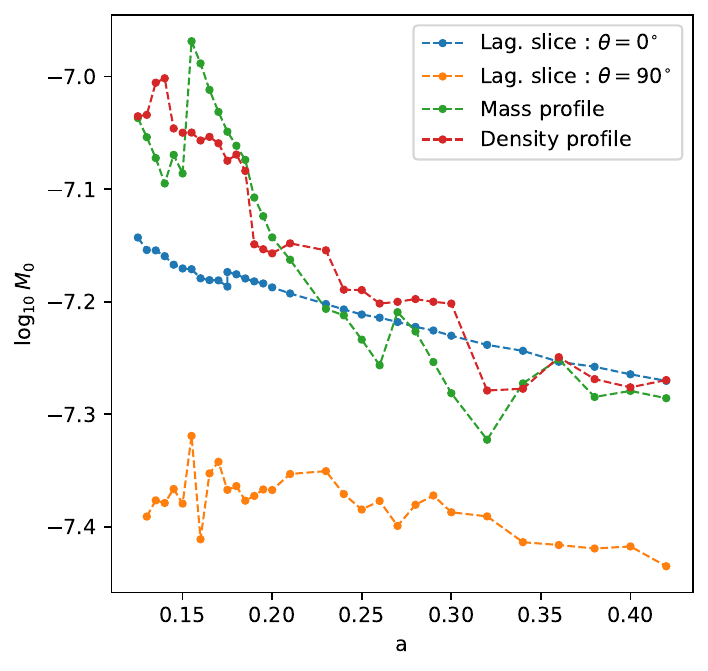}
    \includegraphics[width = 0.33\textwidth]{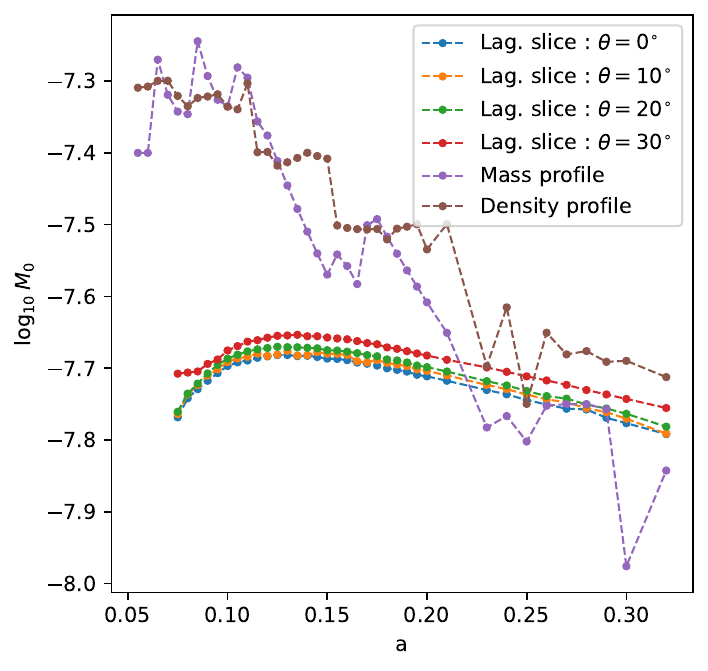}
    \centering
    \caption{Trends in best fit $M_0$ for $r-q$ curves of Lagrangian slices along different angles $\theta$ from the $x$ axis and mass and density profiles for the three cases: Q1D (left), ANI (middle), SYM (right). The parameter $\epsilon$ was fixed at 0.8, 0.7 and 0.5 respectively. }
    \label{fig:M_0_vs_a}
\end{figure*}

\begin{figure*}
    \includegraphics[width = 0.33\textwidth]{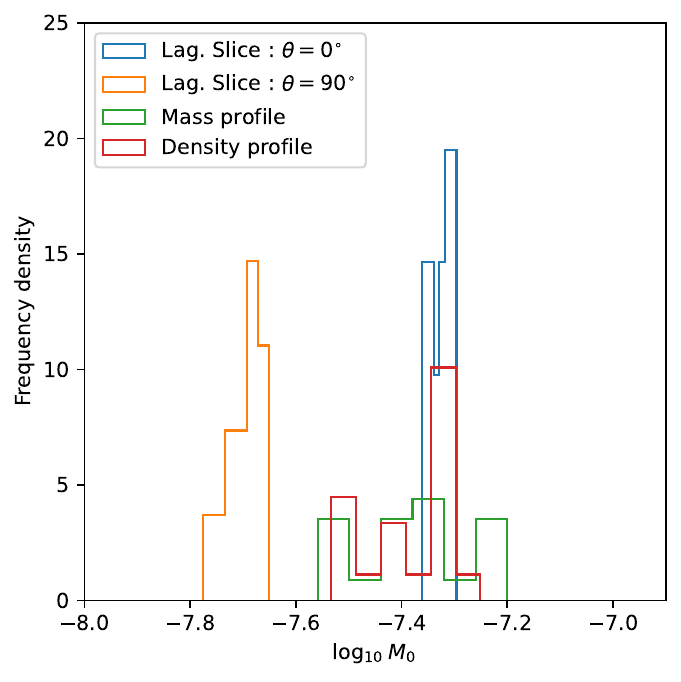}
    \includegraphics[width = 0.33\textwidth]{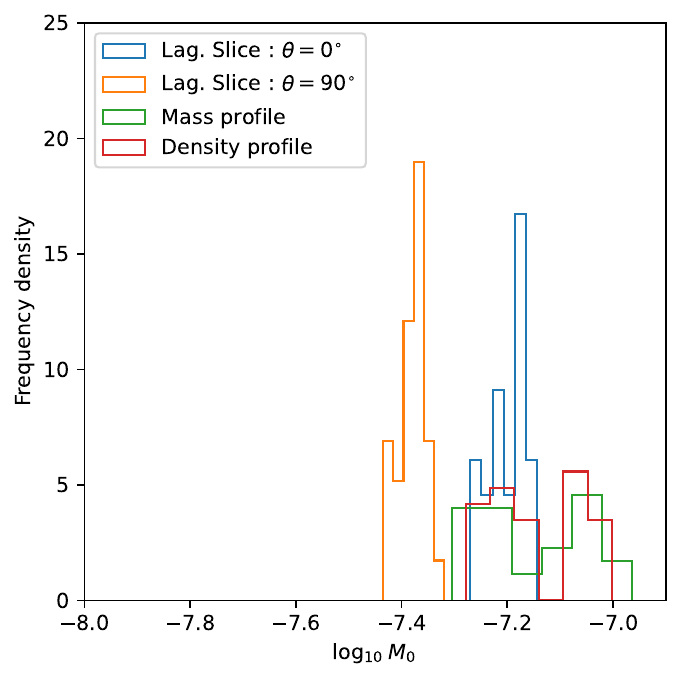}
    \includegraphics[width = 0.33\textwidth]{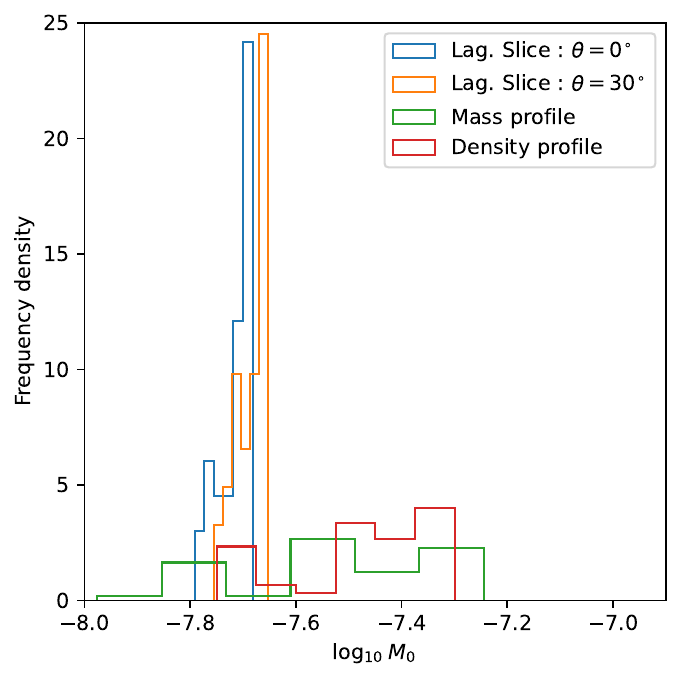}
    
    \centering
    \caption{Distribution of best fit $M_0$ for $r-q$ curves of Lagrangian slices along different angles, mass and density profiles for the three cases: Q1D (left), ANI (middle), SYM (right). The parameter $\epsilon$ was fixed at 0.8, 0.7 and 0.5 respectively.}
    \label{fig:M_0_dist}
\end{figure*}

\begin{figure*}
    \includegraphics[width = 0.33\textwidth]{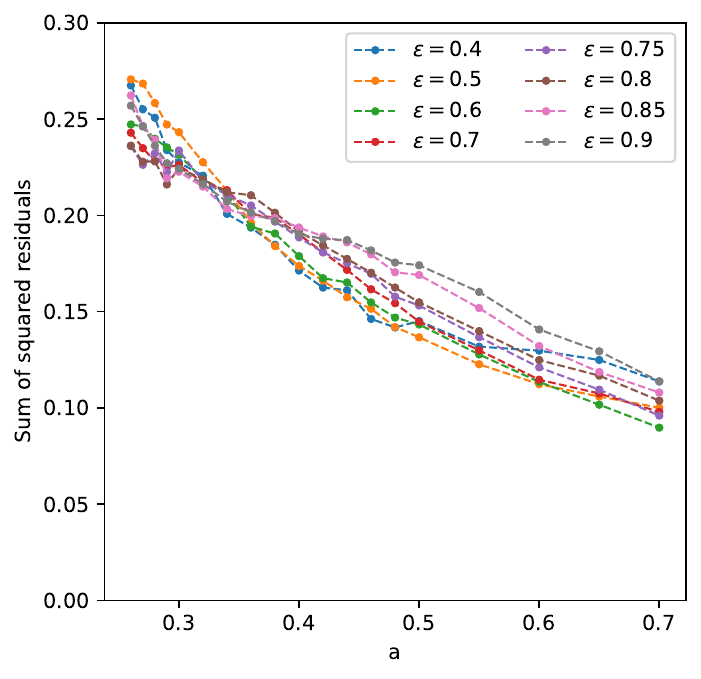}
    \includegraphics[width = 0.33\textwidth]{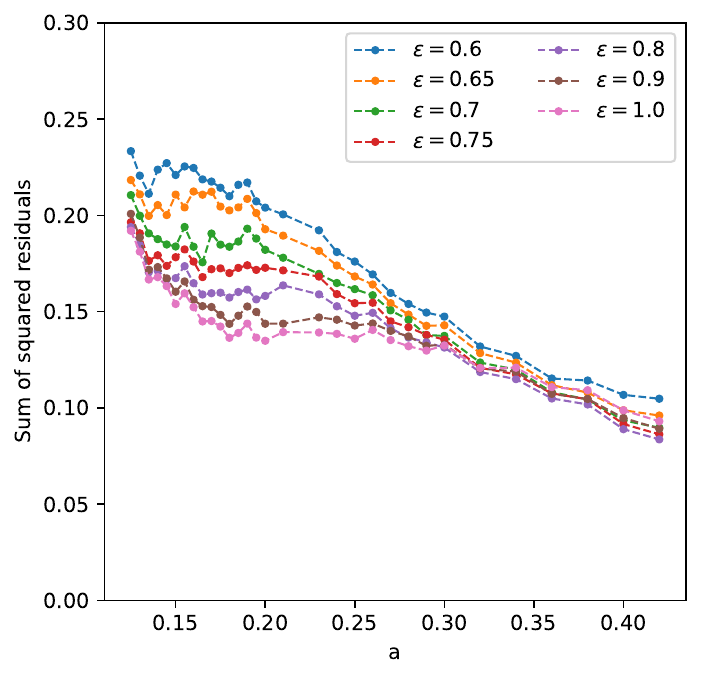}
    \includegraphics[width = 0.33\textwidth]{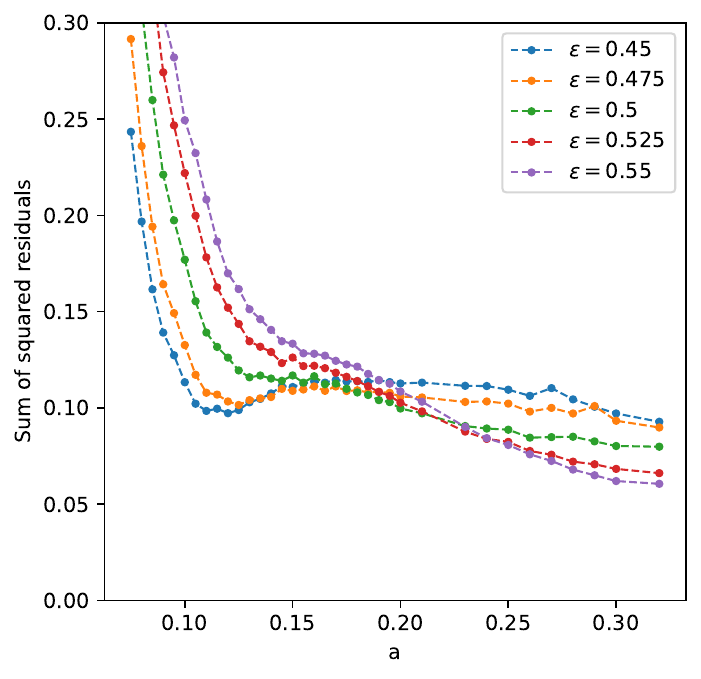}
    
    \centering
    \caption{Trends in residuals $\left( = \frac{1}{N} \Sigma (\Delta r)^2 \right)$ computed using $r-q$ curves of Lagrangian slices along $x$-axis with varying $\epsilon$ for the three cases: Q1D (left), ANI (middle), SYM (right).}
    \label{fig:residual_vs_a}
\end{figure*}

The self-similar fits to the data from simulations were characterised by two parameters: $M_0$ and $\epsilon$, defined in eq. \eqref{eq:init_delta_ss}. Recall that $\epsilon$ characterises the mass accretion rate and frequency of oscillations (refer fig. \ref{fig:self_sim_sol}), whereas $M_0$ sets the scale of the initial perturbation that seeds the halo. As the convergence was slow for a 2-parameter fit, we kept $\epsilon$ fixed and obtained the best fits for $M_0$ using the least-squares method, then repeated the process for several values of $\epsilon$. Through the procedure outlined in sections \ref{subsec:part_traj} and \ref{subsec:mass_rho_profiles}, we have 2 sets of best-fit parameters corresponding to - (i) $r-q$ curves of Lagrangian slices (ii) radial profiles of mass and density - for the three simulations at the available snapshots. We now analyze the trend and spread in their distributions and their variation across the three simulations.

Fig. \ref{fig:M_0_vs_a} shows the trends in the best-fit $M_0$ for $\epsilon$ fixed at $0.8, 0.7$ and $0.5$ for Q1D, ANI and SYM simulations respectively. The best-fit $M_0$ for mass-density profiles is typically higher (for ANI and SYM cases at least) than that for $r-q$ curves at earlier snapshots. Looking at the fits for $r-q$ curves in figure \ref{fig:vl_comp_r_q}, we note that the amplitudes of first 1-2 oscillations in the simulations are higher than that of the self-similar fits, which means that the radii of splashback (outermost caustic) and subsequent few caustics in the simulations are greater than that of their fits. This implies that the initial perturbation $\delta$ in the simulations falls off slower i.e. is broader at larger radii than what the self-similar fits to $r-q$ curves would suggest. In the fits to mass and density profiles (refer fig. \ref{fig:vl_comp_mass_density_profile}) however, the positions of the outer caustics are predominantly matched at earlier snapshots resulting in the best-fit $M_0$ being higher. At later times, when more particles have relaxed to lower amplitudes and more caustics have formed, the weight on matching the first caustics is less and the best-fit $M_0$ decreases.

We expected to see the best-fit $M_0$ for $r-q$ curves of Lagrangian slices along the orthogonal $x, y$ axes to converge for ANI and Q1D cases, suggesting that the dynamics inside the halo turns circularly symmetric with time. But, this is not what we observe. The $r-q$ curves along the $y$-axis are better fit by lower $M_0$ across all the snapshots. It thus seems that the initial conditions do leave their imprints on halo dynamics for 10-11 oscillations about the center at the very least. But since the caustic structure seemed to grow roughly similar at later times (refer fig. \ref{fig:2SIN_vl_snaps}), comparing directly the axial density profiles along orthogonal directions might provide a better demonstration of halos gradually turning circular.

\begin{table}[]
    \centering
    \begin{tabular}{|c|c|c|c|}
    \hline
        & $r-q$   & Mass  & Density \\ \hline
    Q1D & $-7.33 \pm 0.02$ & $-7.37 \pm 0.11$ & $-7.38 \pm 0.09$    \\ \hline
    ANI & $-7.20 \pm 0.03$ & $-7.15 \pm 0.10$  & $-7.14 \pm 0.09$  \\ \hline
    SYM & $-7.72 \pm 0.03$  & $-7.52 \pm 0.18$ & $-7.47 \pm 0.14$   \\ \hline
    \end{tabular}
    \caption{The averages and standard deviations of best-fit $\log_{10} M_0$ from $r-q$ curves along $x$-axis, mass and density profiles for the three simulations.}
    \label{table:ranges_M_0}
\end{table}

Fig. \ref{fig:M_0_dist} shows the histograms of best-fit $M_0$ for $r-q$ curves of Lagrangian slices, mass and density profiles with $\epsilon$ fixed for each simulation. Their average and standard deviations (over snapshots at different times) are tabulated in Table \ref{table:ranges_M_0}. The best-fit $M_0$'s are of the order $M_0\in [10^{-8.0},10^{-6.9}]$ in units of total simulation mass. The relative spread in initial perturbation $\delta$ (refer eq. \eqref{eq:init_delta_ss}) assumed in FG model, corresponding to the spread in best-fit $M_0$ from $r-q$ curves is 12\%, 20\% and 13\% for Q1D, ANI and SYM respectively, which is not unreasonably high given the differences between our simulation setups and the ideal theoretical setup.

To get an idea of the trends in best-fit $\epsilon$ in each simulation, we compare between different values of $\epsilon$, the sum of squared residuals $\left(=\frac{1}{N} \: \Sigma \Delta r^2\right)$ corresponding to the fits of $r-q$ curves of Lagrangian slices along x-axis at each snapshot in Fig \ref{fig:residual_vs_a}. $N$ is the number of data points in each $r-q$ curve and $\Delta r$ is the difference between the position of a Lagrangian point measured from simulation and that expected from the self-similar fit to the curve. The residuals indeed decrease over time implying better fits, which means that the system behaves increasingly in accordance with FG's model of self-similarity. The $\epsilon$ corresponding to the least residue does vary over the span of the simulations: 0.4 - 0.6 for Q1D, 0.8 - 1.0 for ANI and 0.45 - 0.55 for SYM. To demonstrate our fitting procedure and results, we choose the self-similar fits corresponding to $\epsilon = 0.8, 0.7$ and $0.5$ for Q1D, ANI and SYM simulations as representatives. The variation in residuals between different $\epsilon$'s for the Q1D and ANI cases is not as drastic as that for the SYM case. The residuals corresponding to $\epsilon = 0.5$ remain roughly intermediate throughout the SYM simulation. A better choice of $\epsilon$'s, the ones with lower residues, does improve the bounds in fig. \ref{fig:q_bounds} slightly, but the deviations due to relaxation, periodic boundaries and transverse motion are still observable and lead to similar conclusions. The crucial thing to note, at least in Q1D and SYM cases, is that the value of $\epsilon$ which gives the least residue increases with time. This is due to the deficit of infalling matter resulting in a lowering of oscillation frequency with time, as observed in section \ref{subsec:scale_free_space}.
%--------------------------------------------------------------------
\section{Conclusions}
\label{sec:conclusions}

In this work, we aimed to demonstrate the applicability of self-similarity, particularly the FG model, to dark matter halo dynamics using 2D Vlasov simulations of monolithically growing halos seeded by sine wave initial conditions of three different symmetries - Quasi 1D (Q1D), Anisotropic (ANI) and Symmetric (SYM). By testing the FG model against particle trajectories, mass and density profiles, we were able to determine not only the subset of particles that followed the model's predictions at a given instant but also the causes behind the deviations for others. Lastly, the trends and spread of the resulting distribution of best-fit parameters were studied and compared across the three simulations. Below, we summarize our important findings together with the key inferences regarding 3D CDM halos seeded by Gaussian random fields:
\begin{enumerate}
    \item The initial perturbation $\delta$ (refer eq. \eqref{eq:init_delta_ss}) in the simulations certainly does not follow a power law as assumed in FG model. Shell crossing is followed by relaxation which leads to the build-up of a power-law density profile, after which the self-similar fits begin to converge. Particles typically relax after 1-2 oscillations after the last shell-crossing has occurred (which is along $y$-axis in our cases) and then move in agreement with self-similar fits (refer fig. \ref{fig:vl_comp_r_q}). The subset of such particles grows (refer fig. \ref{fig:q_bounds}) as the particles starting farther out gradually relax and the model predictions improve. This feature is common to all three simulations. Therefore, the dynamics of relaxation and formation of prompt cusp in 3D CDM halo simulations cannot be explained by the FG model of self-similarity. 

    \item The particles closer to the boundaries, however, experience additional forces due to the periodic boundaries which slows down their in-fall, causing them to eventually deviate from the self-similar fits. This is purely an artifact of our simulations and does not concern the CDM halos in 3D cosmological simulations.
    
    \item Even after relaxing to self-similar fits, the particle trajectories trace them only for a limited number of oscillations and deviate again. The number of such oscillations is the highest for the SYM simulation. We believe this to be correlated to the varying degrees of transverse motion in our three simulations. Indeed, the SYM case has particle trajectories with the least transverse motion in the inner region of the halo (refer fig. \ref{fig:trans_motion}). Since the FG solutions assume purely radial particle trajectories, it is, but obvious that the SYM simulations appear to be the most 'self-similar' under the FG model. We posit that the particle trajectories deviate once their amplitudes decrease below the radius where the anisotropy parameter transitions from $\beta \sim 1$ to $\beta \sim 0$, where transverse motion starts being significant (refer fig. \ref{fig:beta_param}). The deviation of particle trajectories embedded deep inside the halo interior is therefore a consequence of our choice of a self-similar model. Since CDM halos formed from Gaussian initial conditions are closer to ANI and Q1D cases, we need to consider a more general model that accounts for ellipsoidal collapse and transverse motion of particles while studying CDM halo dynamics in actual cosmologies. Looking at the superposition of particle trajectories at different snapshots in scale-free position-time and phase spaces in fig. \ref{fig:vl_comp_self_sim_space}, we reach to a similar conclusion. Though there exist deviations from the FG model in Q1D and ANI cases, the simulated curves are strikingly self-consistent, which indicates a self-similar pattern more general than the purely radial FG solution alone.

    \item The normalisation and positions of caustics in mass and density profiles are well captured by the self-similar fits (refer fig. \ref{fig:vl_comp_mass_density_profile}). The measured slopes of the central density profiles -0.95, -0.90 and -1.00 in Q1D, ANI and SYM simulations respectively, are also in agreement with FG's prediction of -1 for the asymptotic slope of the density profile, which is independent of the model parameters. However, at late times, the slope at outer radii dips signifying a deficit of infalling mass in the simulations. Though an artifact of our simulations, we speculate that it might actually be relevant for CDM cosmologies with Gaussian initial conditions. Since most of the mass eventually accretes onto some or the other halo, a similar situation of a lack of infalling matter will arise which will cause the self-similar power-law profile to dip towards the outer region of the halo and lead to a running power-law profile which might explain the universal attractor - NFW profile.

    \item Instead of a single set of best-fit parameters $(M_0, \epsilon)$, we obtained a distribution across the snapshots for each simulation (refer figures \ref{fig:M_0_dist} and \ref{fig:residual_vs_a}), which was to be expected since our setups were not as ideal as that assumed in theory. Nevertheless, the spreads in the distributions using particle trajectories were narrow (refer table \ref{table:ranges_M_0}) in the sense that the corresponding error implied on the initial perturbation $\delta$ (refer eq. \eqref{eq:init_delta_ss}), which defines the model parameters, is within reasonable limits. The fact that we could trace 30-60\% of the particle trajectories measured from our simulations within 10\% error for periods spanning over several oscillations after shell-crossing as well as the mass and density profiles using FG self-similar solutions corresponding to a narrow range of parameters justifies its relevance in the study of halo dynamics, at least in the 2D case examined in the present work.

    \item Since the best-fit parameters for particle trajectories of ANI and Q1D along orthogonal directions do not converge within the simulation time (refer fig. \ref{fig:M_0_vs_a}), we could not conclude if the halos gradually turn circular. This suggests that particle trajectories carry the imprints of initial conditions for 11-12 oscillations at the very least. The CDM halos turning increasingly circular is pretty evident from the caustics' structure in the simulations, see Fig.~\ref{fig:2SIN_vl_snaps}. Therefore, looking for the convergence of axial density profiles along orthogonal directions instead might provide better insights in this regard.

    \item Within a particular simulation, the frequency of oscillations of the particles gradually decreases, which we verify by showing that the parameter $\epsilon$ corresponding to the self-similar fit to the particle trajectories with the least residue, gradually increases (refer fig. \ref{fig:residual_vs_a}). Since $\epsilon$ relates inversely to the mass accretion rate, its gradual increase corroborates our claim of a deficit of infalling matter at late times in our simulations.
\end{enumerate}
In conclusion, by performing this exercise of studying 2D Vlasov simulations using the FG self-similar approach, one of the basic models of self-similarity, we could demonstrate self-similar patterns in individual particle trajectories. Though there are deviations, they can be accounted for. The averaged observables like the mass and density profiles are still very well-explained by the model. It helped us gain deeper insights into the signatures of self-similarity to look for in 3D CDM simulations with Gaussian fields. Future work would involve performing similar CDM numerical simulations with sine waves as well as Gaussian initial conditions but in 3D. On the analytical front, we would investigate more general self-similar models which include ellipsoidal collapse \citep[e.g.,][]{Lithwick_2011} and transverse motion of particles and then analyse the particle trajectories, phase-space and mass-density profiles using the procedure established in this work. Therefore, self-similarity could potentially be one of the keys to decoding the complex dynamics in dark matter halos.

%--------------------------------------------------------------------
\begin{acknowledgements}
This work was supported in part by the French Doctoral school ED127, Astronomy and Astrophysics Ile de France (AP), as well as Programme National Cosmology et Galaxies (PNCG) of CNRS/INSU with INP and IN2P3, co-funded by CEA and CNES (SC \& AP). This work was also financially supported by the “PHC Sakura” program (project number: 51203TL, grant number: JPJSBP120243208), implemented by the French Ministry for Europe and Foreign Affairs, the French Ministry of Higher Education and Research and the Japan Society for Promotion of Science (JSPS). This work is additionally supported by the JSPS KAKENHI Grant No. JP23K19050 and No. JP24K17043 (SS); and JP20H05861 and JP21H01081 (AT). Numerical computation with ColDICE was carried out using the HPC resources of CINES (Occigen supercomputer) under the GENCI allocations c2016047568, 2017-A0040407568 and 2018-A0040407568. Post-treatment of ColDICE data was performed on the INFINITY cluster of Institut d’Astrophysique de Paris.
\end{acknowledgements}

\bibliographystyle{aa} % style aa.bst
\bibliography{lib.bib} % your references Yourfile.bib

%\begin{appendix}
%\section{}
%\end{appendix}

\end{document}